\pdfoutput=1
\documentclass[12pt]{article}
\usepackage{amsmath,amssymb,amsbsy,exscale}
\usepackage{graphicx}
\usepackage{epsfig}
\usepackage{multicol}
\usepackage{mathrsfs}
\usepackage{mathtools}
\usepackage{blindtext}
 \usepackage{fancyhdr}
\usepackage{hyperref}
\usepackage{cite}

\usepackage{enumerate}

\usepackage{dsfont}
\usepackage{appendix}

\usepackage{parskip}

\usepackage{xspace}
\usepackage{braket}
\usepackage{array}
\usepackage{feynmp}
\usepackage{graphicx,rotating,colortbl}

\usepackage{pinlabel}
\usepackage[top=1.2in, bottom=1.2in,left=0.8in, right=0.8in,includefoot]{geometry}

\definecolor{nicered}{rgb}{0.7,0.1,0.1}
\definecolor{nicegreen}{rgb}{0.1,0.5,0.1}

\definecolor{rosso}{cmyk}{0,1,1,0.4}
\definecolor{babypink}{rgb}{0.96, 0.76, 0.76}
\definecolor{babyblueeyes}{rgb}{0.63, 0.79, 0.95}
\definecolor{azure(colorwheel)}{rgb}{0.0, 0.5, 1.0}
\definecolor{amethyst}{rgb}{0.6, 0.4, 0.8}
\definecolor{MyDarkBlue}{rgb}{0,0.1,0.7}

\definecolor{secnum}{RGB}{13,151,225}
\definecolor{ptcbackground}{RGB}{212,237,252}
\definecolor{ptctitle}{RGB}{0,177,235}
\definecolor{blus}{cmyk}{1,1,0,0.1}
\definecolor{verdes}{cmyk}{0.99,0,0.59,0.65}
\definecolor{rossos}{cmyk}{0,1,1,0.55}
\definecolor{redy}{cmyk}{0,1,1,0.7}
\definecolor{greeny}{cmyk}{0.99,0,0.59,0.98}
\definecolor{green-go}{cmyk}{0.79,0,0.59,0.5}

\usepackage{hyperref, pdfsync, epstopdf, enumerate}
\hypersetup{colorlinks,bookmarksopen,bookmarksnumbered,citecolor={nicegreen},
linkcolor={MyDarkBlue},pdfstartview=FitH,urlcolor={amethyst}}
\usepackage{amsfonts}
\usepackage{slashed}

\usepackage{verbatim}
\usepackage{doi}
\usepackage{url}
\usepackage{array}

\usepackage{cite}
\usepackage{float}
\usepackage[labelfont=bf]{caption}
\usepackage{subcaption}
\usepackage{parskip}
\usepackage{hhline}

\numberwithin{equation}{section}

\usepackage{sectsty}

\usepackage{mdframed}
\usepackage{titletoc}

\titlecontents{lsection}
  [5.8em]{\sffamily}
  {\color{greeny}\contentslabel{2.3em}\normalcolor}{}
  {\titlerule*[1000pc]{.}\contentspage\\\hspace*{-5.8em}\vspace*{5pt}%
    \color{white}\rule{\dimexpr\textwidth-15.5pt\relax}{1pt}}

\usepackage{titlesec}
\titlelabel{\thetitle.\quad}

\newcommand{\tmtextbf}[1]{{\bfseries{#1}}}
\newcommand{\tmtextrm}[1]{{\rmfamily{#1}}}
\newcommand{\gappeq}{{\rlap{{\raise}.5ex\text{\ensuremath{>}}}{{\lower}.5ex\text{\ensuremath{\sim}}}}}
\newcommand{\lappeq}{{\rlap{{\raise}.5ex\text{\ensuremath{<}}}{{\lower}.5ex\text{\ensuremath{\sim}}}}}

\DeclareMathOperator{\tr}{tr}
\newcommand{\I}{\tmtextrm{1{\kern}-.24em l}}

\newcommand{\newc}{\newcommand}
\newc{\be}{\begin{equation}}
\newc{\ee}{\end{equation}}
\newc{\bal}{\begin{align}}
\newc{\eal}{\end{align}}
\newc{\ba}{\begin{eqnarray}}
\newc{\ea}{\end{eqnarray}}
\newc{\bea}{\begin{eqnarray*}}
\newc{\eea}{\end{eqnarray*}}
\newc{\D}{\partial}
\newc{\som}{\sin\omega}
\newc{\com}{\cos\omega}
\newc{\sth}{\sin\theta}
\newc{\cth}{\cos\theta}
\newc{\stom}{\sin^2\omega}
\newc{\ctom}{\cos^2\omega}
\newc{\stth}{\sin^2\theta}
\newc{\ctth}{\cos^2\theta}
\newc{\ie}{{\it i.e.} }
\newc{\eg}{{\it e.g.} }
\newc{\etc}{{\it etc.} }
\newc{\etal}{{\it et al.}}

\newcommand{\GeV}{\,\,\mathrm{GeV}}

\newcommand{\lapproxeq}{\lower .7ex\hbox{$\;\stackrel{\textstyle
<}{\sim}\;$}}
\newcommand{\gapproxeq}{\lower .7ex\hbox{$\;\stackrel{\textstyle
>}{\sim}\;$}}
\newcommand{\stackdown}[2]{\lower 1.4ex\hbox{$\;\stackrel{\textstyle{#1}}
{\scriptstyle{#2}}\;$}}

\makeatletter
\def\@xfootnote[#1]{%
  \protected@xdef\@thefnmark{#1}%
  \@footnotemark\@footnotetext}
\makeatother

\numberwithin{equation}{section}

\begin{document}

\topmargin -1.0cm
\oddsidemargin -0.5cm
\evensidemargin -0.5cm

%IFT preprint number
%\vspace{-1cm}
% \hspace{12cm}IFT-UAM/CSIC-15-140
 
{\vspace{1cm}}
\begin{center}
\vspace{1cm}

 {\huge  \tmtextbf{ 
{Dark Matter from a Classically Scale-Invariant $\bf{SU(3)_X}$  }}} {\vspace{.5cm}}\\

\vspace{1.4cm}

{\large  {\bf Alexandros Karam$^{1}$\footnote[*]{email: {\href{mailto:alkaram@cc.uoi.gr}{alkaram@cc.uoi.gr}}} and Kyriakos Tamvakis$^{1,2}$\footnote[$\dagger$]{email: {\href{mailto:tamvakis@uoi.gr}{tamvakis@uoi.gr}}}
}
%\vspace{.4cm}\\
%{\large }\\
\vspace{.3cm}

{\it }\

{\em  \normalsize 

$^1$~University of Ioannina, Physics Department, Section of Theoretical Physics, GR--45110 Ioannina, Greece\\ 

\vspace{0.2cm}

$^2$~CERN, Theoretical Physics Department, Geneva 23, CH-1211, Switzerland}

\vspace{0.5cm}

}
\vspace{1.2cm}
 \end{center}
\noindent --------------------------------------------------------------------------------------------------------------------------------

\vspace{-0.3cm}

% \begin{abstract}
\begin{center}
{\large \bf  Abstract}
\end{center}

\noindent {\large  In this work we study a classically scale-invariant extension of the Standard Model in which the dark matter and electroweak scales are generated through the Coleman-Weinberg mechanism. The extra $SU(3)_X$ gauge factor gets completely broken by the vacuum expectation values of two scalar triplets. Out of the eight resulting massive vector bosons the three lightest are stable due to an intrinsic $Z_2\times Z_2'$ discrete symmetry and can constitute dark matter candidates. We analyze the phenomenological viability of the predicted multi-Higgs sector imposing theoretical and experimental constraints. We perform a comprehensive analysis of the dark matter predictions of the model solving numerically the set of coupled Boltzmann equations involving all relevant dark matter processes and explore the direct detection prospects of the dark matter candidates.}

%\end{abstract}

\vspace{.9cm}

\noindent --------------------------------------------------------------------------------------------------------------------------------

%\vspace{1.1cm}

{\sc Keywords: } {\small Classical Scale Invariance, Beyond Standard Model, Higgs boson, Dark Matter}

\newpage

{
\tableofcontents
}

\vspace{.9cm}
\noindent --------------------------------------------------------------------------------------------------------------------------------
%%%%%%%%%%%%%%%%%%%%%%%%%%%%%%%%%
\section{Introduction}
\label{intro}

The first run of the LHC culminated with the discovery~\cite{Aad2012,Chatrchyan2012} of the 125 GeV Higgs boson~\cite{Englert1964,Higgs1964,Higgs1964a,Guralnik1964}. The Standard Model (SM) is now complete and has successfully passed every experimental test. Nevertheless, it comes short of describing various phenomena such as the nature of dark matter, the nonzero neutrino masses, the asymmetry between matter and antimatter. It also cannot explain the origin of the electroweak scale and why strong interactions seem to preserve the $CP$ symmetry. A more fundamental theory should be able to address these issues and also accommodate a particle physics description of cosmological inflation. The second run of the LHC is now underway and will hopefully provide us with solutions to some of these problems and point us to a direction for physics beyond the Standard Model.

In the SM, the Higgs field $H$ enters the Lagrangian through the scalar potential 
%%%%%%%%%%%%%%%%%%%%%%%%%%%
\be 
V(H) = - m^2 H^\dagger H + \lambda_h \left( H^\dagger H \right)^2,
\ee
%%%%%%%%%%%%%%%%%%%%%%%%%%%
where $\lambda_h$ is the Higgs self-coupling and $m^2 > 0$ is the mass parameter responsible for spontaneously breaking the electroweak symmetry. The latter is the only dimensionful parameter in the SM and its quadratic sensitivity with respect to higher scales is what causes the \textit{hierarchy problem}. Setting $m^2 = 0$ results in a manifestly classically scale-invariant (CSI) theory~\cite{Bardeen1995}. In 1973 Coleman and E. Weinberg (CW)~\cite{Coleman1973} considered scalar QED and showed that classical scale symmetry gets broken at the quantum level due to logarithmic corrections and that the gauge symmetry breaking scale can arise through \textit{dimensional transmutation}. Three years later Gildener and S. Weinberg (GW)~\cite{Gildener1976} generalized their mechanism by considering an arbitrary number of scalar fields. However, an implementation of the CW mechanism in the SM is not phenomenologically viable due to the large top mass that renders the effective potential unstable. This situation can be remedied by extending the SM with new scalar and/or vector degrees of freedom which contribute positively to the effective potential.

The measured value of the Higgs boson mass $M_h = 125.09 \pm 0.24$ GeV~\cite{Aad2015} gives $\lambda(M_t)\approx 0.1285$~\cite{Sirlin1986, Holthausen2012} at the scale of the top mass. Because of the large contribution of the top Yukawa coupling in its renormalization group equation (RGE), $\lambda_h$ runs negative above scales of $\mathcal{O}(10^{10}\,\mathrm{GeV})$ which results in the vacuum being in a metastable state~\cite{Degrassi2012, Buttazzo2013, Zoller2014, Branchina2013, Branchina2014, Andreassen2014, Branchina2015, Bezrukov:2014ipa, Espinosa2015b, Bednyakov2015a}. In order to (fully) stabilize the potential, one needs to couple the Higgs field with extra bosonic fields that contribute positively to the RGE of $\lambda_h$.

A classically scale invariant extension of the SM can in principle solve both the hierarchy and the vacuum stability problems. Various CSI models have recently been proposed. The authors of~\cite{Dias2006a, Meissner2007, Foot2007, Foot2007a, Foot2008, Foot2010, Alexander-Nunneley2010, Foot2011a, Foot2011, Ishiwata2012, Lee2012, Farzinnia2013, Chway2014a, Lindner2014, Davoudiasl2014, Gabrielli2014, Antipin2014, Guo2015a, Farzinnia2014, Kannike2014, Kang2015a, Humbert2015, Endo2015b, Sannino2015a, Kang2015b, Endo2015c, Farzinnia2015a, Endo2016, Ahriche2016, Wang2015, Ghorbani2016, Farzinnia2016, Helmboldt2016, Ferreira2016, Ahriche2016a, Ahriche2016b, Kawana2016, Wu2016, Marzola2016} extended only the scalar sector, while the authors of~\cite{Hempfling1996, Chang2007, Iso2009, Iso2009a, Holthausen2010, Iso2013, Englert2013, Chun2013a, Chivukula2013a, Khoze2013, Khoze2013a, Hambye2013, Carone2013, Heikinheimo2014, Holthausen2013, Hashimoto2014, Benic2014, Khoze2014c, Benic2015a, Kubo2014b, Pelaggi2015, Guo2015b, Haba2016, Orikasa2015, Oda2015b, Haba2015a, Ametani2015a, Carone2015a, Altmannshofer2015, Plascencia2015a, Karam2015, Wang2015a, Das2015, Haba2016a, Okada2015a, Karam:2016bhq, Jinno2016, Das2016, Khoze2016, Kannike2016a, Hatanaka2016} extended the gauge sector as well with Abelian or non-Abelian gauge symmetries. Some of these models have the appealing feature that they also predict stable and weakly interacting massive particles (WIMPs) which can be viable candidates for dark matter (DM).

In this work, we propose a CSI extension of the SM where a new $SU(3)_X$ gauge symmetry can provide massive gauge fields that can account for the observed DM relic density. The hidden sector will be broken completely by two scalar triplets. These will have portal couplings with the Higgs field and will help in the stabilization of the potential. The scalar sector will consist of three Higgs-like particles, one of which will be massless at tree level but will nevertheless acquire a nonzero mass once we consider the full one-loop scalar potential. All eight of the extra gauge bosons will become massive, while the three lightest will be stable due to their parities under an intrinsic $Z_2 \times Z'_2$ discrete symmetry of $SU(3)_X$. These three dark gauge bosons will be our DM candidates. Because of the rich structure of the extra gauge group, the computation of the DM relic density will include various types of processes apart from DM annihilations, such as semiannihilations, coannihilations, and DM conversions.

The layout of the paper is the following. In the next section we present the model and calculate the masses of the new fields. In Sec. \ref{sec:pheno} we impose various theoretical and experimental constraints on the model. Then, in Sec. \ref{sec:DM} we give a detailed analysis of the system of Boltzmann equations that need to be solved in order to obtain the DM relic abundance, and we also focus on the role of coannihilations and DM conversion processes. Furthermore, we examine the direct detection prospects of the DM candidates. Finally, we summarize and conclude in Sec. \ref{sec:conclusions}. Useful formulas are presented in Appendices \ref{app:obliques}, \ref{app:RGEs}, and \ref{app:kin}.

%%%%%%%%%%%%%%%%%%%%%%%%%%%%%
\section{The Model}
\label{sec:model} 
%%%%%%%%%%%%%%%%%%%%%%%%%%%%%
We begin with a CSI version of the Standard Model and consider an $SU(3)_X$ extension of its gauge symmetry in order to accommodate the presence of dark matter. The non-CSI version of this model was recently considered in Ref.~\cite{Gross2015}. The breaking of the gauge symmetry $SU(3)_C\times SU(2)_L\times U(1)_Y\times SU(3)_X\rightarrow\,SU(3)_C\times U(1)_{em}$ is achieved through the Coleman-Weinberg mechanism~\cite{Coleman1973}. In addition to the new $SU(3)_X$ gauge bosons, referred to as ``dark" gauge bosons, the model contains a pair of complex scalars $\Phi_1(1,1,0;3)$ and $\Phi_2(1,1,0;3)$ transforming as singlets under the Standard Model gauge group and as triplets under $SU(3)_X$, referred to as ``dark" scalars. In this section we explore the scalar and gauge sectors of the model. First, we present the tree-level potential. Employing the Gildener-Weinberg formalism~\cite{Gildener1976}, we minimize the tree-level potential at a definite energy scale which defines a flat direction among the scalar fields. Then, we compute the tree-level scalar and dark gauge boson masses. One of the scalar bosons turns out to be massless at tree level and corresponds to the pseudo--Nambu-Goldstone boson (pNGB) of broken scale symmetry. Finally, we present the one-loop effective potential which becomes dominant along the flat direction and greatly lifts the mass of the pNGB.
%%%%%%%%%%%%%%%%%%%%%%%%%%%%%
\subsection{Tree-level potential}
%%%%%%%%%%%%%%%%%%%%%%%%%%%%%
The most general renormalizable and scale-invariant tree-level scalar potential involving the standard Higgs doublet $H$ and the dark triplets $\Phi_1,\,\Phi_2$ is
\be
\begin{split}
V_0\,&=\,\lambda_h ( H^{\dagger} H )^2\,+\,\lambda_1 ( \Phi_1^{\dagger} \Phi_1 )^2 + \lambda_2 ( \Phi_2^{\dagger} \Phi_2 )^2 - \lambda_3 ( \Phi_1^{\dagger} \Phi_1 ) ( \Phi_2^{\dagger} \Phi_2 ) + \lambda_4 ( \Phi_1^{\dagger} \Phi_2 ) ( \Phi_2^{\dagger} \Phi_1 ) \\
 &+ \left[ \frac{\lambda_5}{2}(\Phi_1^{\dagger}\Phi_2)^2 + \lambda_6 ( \Phi_1^{\dagger} \Phi_1 )(\Phi_1^{\dagger}\Phi_2) +\lambda_7 ( \Phi_2^{\dagger} \Phi_2 )(\Phi_1^{\dagger}\Phi_2)+ \mathrm{H.c.} \right]-\lambda_{h1} ( H^{\dagger} H ) ( \Phi_1^{\dagger} \Phi_1 ) \\
& +\,\, \lambda_{h2} ( H^{\dagger} H ) ( \Phi_2^{\dagger} \Phi_2 )\,\, - \left( \lambda_{h12} ( H^{\dagger} H ) (\Phi_1^{\dagger}\Phi_2) + \mathrm{H.c.} \right) , 
\end{split}
\ee
%%%%%%%%%%%%%%%%%%%%%%%%%%%%
where all appearing coupling constants are taken to be real and positive. Notice that we have assumed negative signs for the $\lambda_{h1}$ and $\lambda_3$ portal couplings as the basic seed of symmetry breaking. Out of the $12$ degrees of freedom included in $\Phi_1$, $\Phi_2$, $8$ are Higgsed away. Using gauge freedom and removing $5$ of them from $\Phi_1$ and $3$ from $\Phi_2$, we end up in the unitary gauge with $\Phi_1$ containing $1$ and $\Phi_2$ $3$ real degrees of freedom 
%%%%%%%%%%%%%%%%%%%%%%%%%%%%
\be 
\Phi_1\,=\,\frac{1}{\sqrt{2}}\left(\begin{array}{c}
0\\
\,\\
0\\
\,\\
v_1+\phi_1
\end{array}\right)\,,\qquad 
\Phi_2\,=\,\frac{1}{\sqrt{2}}\left(\begin{array}{c}
0\\
\,\\
v_2+\phi_2\\
\,\\
(v_3+\phi_3)+i(v_4+\phi_4)
\end{array}\right).
\ee
%%%%%%%%%%%%%%%%%%%%%%%%%%%%
Assuming $CP$ invariance implies that all vacuum expectation values (VEVs) are real and $v_4=0$. The extra $SU(3)_X$ can be completely broken if at least two of the remaining VEVs are nonzero, so we further assume $v_3=0$ for simplicity. The standard Higgs will correspond to $1$ real degree of freedom
%%%%%%%%%%%%%%%%%%%%%%%%%%%%
\be 
H\,=\,\frac{1}{\sqrt{2}}\left(\begin{array}{c}
0\\
\,\\
v_h+h
\end{array}\right)\,.
\ee
%%%%%%%%%%%%%%%%%%%%%%%%%%%
The scalar potential is further simplified if we impose invariance of the potential under the discrete symmetry 
%%%%%%%%%%%%%%%%%%%%%%%%%%%
\be 
\Phi_2 \rightarrow \,-\Phi_2,
\ee
%%%%%%%%%%%%%%%%%%%%%%%%%%%
which implies
%%%%%%%%%%%%%%%%%%%%%%%%%%%
\be 
\lambda_6=\lambda_7=\lambda_{h12}=0. 
\ee
%%%%%%%%%%%%%%%%%%%%%%%%%%%
Omitting the VEVs for the moment, the resulting potential is
%%%%%%%%%%%%%%%%%%%%%%%%%%%
\ba
V_0\,&=&\,\frac{\lambda_h}{4}h^4+\frac{\lambda_1}{4}\phi_1^4+\frac{\lambda_2}{4}\phi_2^4-\frac{\lambda_{h1}}{4}h^2\phi_1^2+\frac{\lambda_{h2}}{4}h^2\phi_2^2-\frac{\lambda_3}{4}\phi_1^2\phi_2^2 \nonumber \\
&&+\frac{\lambda_2}{4}(\phi_3^2+\phi_4^2)^2+\left(\frac{\lambda_2}{2}\phi_2^2+\frac{\lambda_3}{4}\phi_1^2+\frac{\lambda_4}{4}\phi_1^2+\frac{\lambda_{h2}}{4}h^2\right)\left(\phi_3^2+\phi_4^2\right)+\frac{\lambda_5}{4}\phi_1^2\left(\phi_3^2-\phi_4^2\right).
\ea
%%%%%%%%%%%%%%%%%%%%%%%%%%%
The above potential is bounded from below if the following conditions~\cite{Kannike2012,Chakrabortty2014,Kannike2016} are satisfied for all energies up to  the Planck scale\footnote{In fact, a more rigorous treatment shows that we must replace $\lambda_3$ with $\lambda_3 + \min\left[ 0, \lambda_4 + \lambda_5, \lambda_4 - \lambda_5 \right]$ in the stability conditions \eqref{stabilityconditions1}--\eqref{stabilityconditions3}. However, we shall assume $\lambda_4 + \lambda_5 > 0$ and $\lambda_4 - \lambda_5 > 0$, resulting in positive masses for the fields $\phi_3$ and $\phi_4$ [cf. \eqref{phi3mass}--\eqref{phi4mass}]. Therefore $\min\left[ 0, \lambda_4 + \lambda_5, \lambda_4 - \lambda_5 \right]=0$.}:
%%%%%%%%%%%%%%%%%%%%%%%
\be
\lambda_h\,\geq\,0,\,\lambda_{1}\,\geq\,0,\,\lambda_{2}\,\geq\, 0, \label{stabilityconditions1}
\ee
%%%%%%%%%%%%%%%%%%%%%%%
\be
2\sqrt{\lambda_h\lambda_1} - \lambda_{h 1} \geq 0, \quad 2\sqrt{\lambda_h\lambda_2} + \lambda_{h 2} \geq 0, \quad 2\sqrt{\lambda_1\lambda_2} - \lambda_3 \geq 0, \label{stabilityconditions2}
\ee
%%%%%%%%%%%%%%%%%%%%%%%
\be
4\lambda_h\lambda_1\lambda_2 - \left(\lambda_{h 1}^2\lambda_2+\lambda_{h 2}^2\lambda_1 + \lambda_3^2\lambda_h\right) + \lambda_{h 1}\lambda_{h 2}{\lambda_3}\,\geq\,0. \label{stabilityconditions3}
\ee
%%%%%%%%%%%%%%%%%%%%%%

%%%%%%%%%%%%%%%%%%%%%%%%%%%
\subsection{Scalar masses}
%%%%%%%%%%%%%%%%%%%%%%%
%%%%%%%%%%%%%%%%%%%%%%%
Gauge symmetry breaking to $SU(3)_C\times U(1)_{em}$ can arise through the nonzero VEVs $v_h,\,v_1,\,v_2$. Since the tree-level potential does not contain any dimensionful parameters, this can only occur via the Coleman-Weinberg mechanism~\cite{Coleman1973}. Having multiple scalars, we will make use of the Gildener-Weinberg approach~\cite{Gildener1976} in order to minimize the potential. The tree-level potential is minimized at a particular renormalization scale $\mu = \Lambda$ which defines the flat direction among the VEVs. The corresponding equations read~\cite{Gildener1976}
%%%%%%%%%%%%%%%%%%%%%%%%%%
\ba
&&\lambda_h \left( \Lambda \right) v_h^4+\lambda_1\left( \Lambda \right)v_1^4+\lambda_2\left( \Lambda \right)v_2^4-\lambda_3\left( \Lambda \right)v_1^2v_2^2-\lambda_{h1}\left( \Lambda \right)v_h^2v_1^2+\lambda_{h2}\left( \Lambda \right)v_h^2v_2^2=0, \label{mincond1}\\
&&2\lambda_h\left( \Lambda \right) v_h^2-\lambda_{h1}\left( \Lambda \right)v_1^2+\lambda_{h2}\left( \Lambda \right)v_2^2\,=\,0, \label{mincond2}\\
&&2\lambda_1\left( \Lambda \right)v_1^2-\lambda_3\left( \Lambda \right)v_2^2-\lambda_{h1}\left( \Lambda \right)v_h^2\,=\,0, \label{mincond3}\\
&&2\lambda_2\left( \Lambda \right)v_2^2-\lambda_3\left( \Lambda \right)v_1^2+\lambda_{h2}\left( \Lambda \right)v_h^2\,=\,0. \label{mincond4}
\ea
%%%%%%%%%%%%%%%%%%%%%%%%%%
Along the flat direction, the shifted scalar fields may be written as
%%%%%%%%%%%%%%%%%%%%%%%%%% 
\be
h\,=\,(\varphi+v)\,n_h, \quad \phi_1\,=\,(\varphi+v)\,n_1, \quad \phi_2\,=\,(\varphi+v)\,n_2,
\ee
%%%%%%%%%%%%%%%%%%%%%%%%%%
where $\varphi^2 = h^2 + \phi^2_1 + \phi^2_2 $ and the overall VEV $v$ is $v^2 = v^2_h + v^2_1 + v^2_2$, with $n_h^2 + n_1^2 + n_2^2 = 1$.

The mass matrix of the three scalar fields that participate in the symmetry breaking can be read off from the shifted tree-level potential to be
%%%%%%%%%%%%%%%%%%%%%%%%%%
\be 
{\mathcal{M}}_0^2\,=\,v^2\,\left(\begin{array}{ccc}
2\lambda_hn_h^2\,&\,-n_hn_1\lambda_{h1}\,&\,n_hn_2\lambda_{h2}\\
\,&\,&\,\\
-n_hn_1\lambda_{h1}\,&\,2\lambda_1n_1^2\,&\,-n_1n_2\lambda_3\\
\,&\,&\,\\
n_hn_2\lambda_{h2}\,&\,-n_1n_2\lambda_3\,&\,2\lambda_2n_2^2
\end{array}\right){\label{MASS}}
\ee
%%%%%%%%%%%%%%%%%%%%%%%%%%
in the $(h,\,\phi_1,\,\phi_2)$ basis. Next, we may consider a general rotation
\be 
{\mathcal{R}}\,{\mathcal{M}}_0^2\,{\mathcal{R}}^{-1}\,=\,{\mathcal{M}}_{d}^2\,\,\,\,\Longrightarrow\,\left(\begin{array}{c}
h \\
\phi_1 \\
\phi_2 
\end{array}
\right)
=  
\mathcal{R}^{-1}
\left(
\begin{array}{c}
h_1 \\
h_2 \\
h_3
\end{array}
\right),
\ee
%%%%%%%%%%%%%%%%%%%%%%%%
in terms of the rotation matrix ${\mathcal{R}}^{-1}$ given by
%%%%%%%%%%%%%%%%%%%%%%%%
\be 
{\cal{R}}^{-1}\,=\,\left( \begin{array}{ccc}
\cos\alpha\cos\beta  & \sin\alpha & \cos\alpha\sin\beta  \\
-\cos\beta\cos\gamma\sin\alpha + \sin\beta\sin\gamma      & \cos\alpha\cos\gamma   & -\cos\gamma\sin\alpha\sin\beta - \cos\beta\sin\gamma \\
-\cos\gamma\sin\beta - \cos\beta\sin\alpha\sin\gamma  & \cos\alpha\sin\gamma & \cos\beta\cos\gamma - \sin\alpha\sin\beta\sin\gamma
\end{array}
\right).
\label{rotationmatrix}
\ee
Two of these rotation angles may be chosen to be related to the flat direction through
\be 
\begin{array}{l}
n_h\,=\,\sin\alpha ,\\
n_1\,=\,\cos\alpha\cos\gamma ,\\
n_2\,=\,\cos\alpha\sin\gamma .
\end{array}
\label{vevs2}
\ee
Then, ${\cal{M}}_d^2$ is diagonal, provided that the following relation is satisfied:
\be
\tan2\beta = \frac{v_h v_1 v_2 v \left( \lambda_{h2}+\lambda_{h1} \right)}{\left(\lambda_1+\lambda_2+\lambda_3\right) v_1^2 v_2^2 - \lambda_h v_h^2 v^2} \,\, .
\ee
The resulting tree-level masses include a zero eigenvalue, namely, $M_{h_2}\,=\,0$, which corresponds to the pNGB of broken scale invariance. Of course, this mass will be strongly lifted at the one-loop level. The other two eigenvalues $M_{h_1},\,M_{h_3}$ are given by complicated expressions in terms of the overall VEV, the angles, and the scalar couplings. In addition to the above three scalar states there are also the scalar fields $\phi_3,\,\phi_4$, which we did not include in the above analysis. These fields do not receive a VEV but obtain tree-level masses as soon as the gauge symmetry breaking is established. 
As we will see in Sec. \ref{OneLoopPot}, radiative corrections will strongly affect only the flat direction defined by $h_2$, while the masses of $\phi_3,\,\phi_4,\,h_1,\,h_3$ will stay close to their tree-level values.
%%%%%%%%%%%%%%%%%%%%%%%%%%%%%
%%%%%%%%%%%%%%%%%%%%%%%%%%
%%%%%%%%%%%%%%%%%%%%%%%%%%
\subsection{Dark gauge boson masses}
%%%%%%%%%%%%%%%%%%%%%%%%%%%%%
%%%%%%%%%%%%%%%%%%%%%%%%%%%%%
The $SU(3)_X$ gauge fields enter the Lagrangian through the kinetic terms
%%%%%%%%%%%%%%%%%%%%%%%%%%%%%
\be 
\mathcal{L}_X = -\frac{1}{2} \tr \{ X_{\mu\nu} X^{\mu\nu} \} + \lvert D_\mu \Phi_1 \rvert^2 + \lvert D_\mu \Phi_2 \rvert^2 ,
\ee
%%%%%%%%%%%%%%%%%%%%%%%%%%%%%
where the field strength tensor is defined as $X_{\mu\nu} = \D_\mu X_\nu - \D_\nu X_\mu + i g_X \left[ X_\mu, X_\nu \right]$ and the covariant derivative of $\Phi_i$ has the form $D_\mu \Phi_i = \D_\mu \Phi_i + i g_X X_\mu \Phi_i$.

Following Ref.~\cite{Gross2015}, we consider the discrete symmetry $Z_2 \times Z'_2$ of the $SU(3)$ generators in the Gell-Mann basis, where the first $Z_2$ corresponds to a gauge transformation, while the second $Z'_2$ is identified with complex conjugation. The parities of the gauge fields $X_\mu$ and the scalar fields $\Phi_i$ under $Z_2 \times Z'_2$ are summarized in Table \ref{tab:parities}. This discrete symmetry is important for the identification of dark matter since the lightest fields with nontrivial discrete signatures will not be able to decay to Standard Model matter.

%%%%%%%%%%%%%%%%%%%%%%%%%%%%%%%%%%%
%%%%%%%%%%%%%%%%%%%%%%%

\begin{table}[H]
\centering
\begin{tabular}{ | c | c | }
\hline 
\rowcolor[cmyk]{0.2,0.1,0.1,0.2} Fields & $Z_2 \times Z'_2$ \\ 
%\hline \hline
\hhline{|=|=|}
\rowcolor[cmyk]{0,0,0.2,0} $h, \, \phi_1, \, \phi_2, \, \phi_3, \, X^7_\mu $ & $\left( + , + \right)$ \\ 
%\hline
\rowcolor[cmyk]{0,0.2,0,0.1} $X^2_\mu, \, X^5_\mu$ & $\left( - , + \right)$\\
%\hline 
\rowcolor[cmyk]{0.1,0,0.1,0} $X^1_\mu, \, X^4_\mu$ & $\left( - , - \right)$\\
%\hline  
\rowcolor[cmyk]{0.2,0.0,0,0.0} $\phi_4, \, X^3_\mu, \, X^6_\mu, \, X^8_\mu$ & $\left( + , - \right)$\\
\hline 
\end{tabular}
\caption{Gauge and scalar fields parities under $Z_2 \times Z'_2$.}
\label{tab:parities}
\end{table}
%%%%%%%%%%%%%%%%%%%%%%%
%%%%%%%%%%%%%%%%%%%%%%%%%%%%%%%%%%%
For the particular choice of nonzero $v_{1,2}$ and $v_{3,4} = 0$, there is only one mixing term, $X_\mu^3 X^{\mu 8}$, among the dark gauge fields. The gauge boson mass matrix has the form 

%%%%%%%%%%%%%%%%%%%%%%%%%%%%%%%%%%%
%%%%%%%%%%%%%%%%%%%%%%%%%%%%%%%%%%%
\be 
\mathcal{M}^2_X = \frac{g^2_X}{4} \left( 
\begin{array}{cccccccc}
v^2_2&  0  &           0           &  0  &  0  &     0     &     0     &    0  \\
  0  &v^2_2&           0           &  0  &  0  &     0     &     0     &    0  \\
  0  &  0  &        v^2_2          &  0  &  0  &     0     &     0     & -\frac{v^2_2}{\sqrt{3}} \\
  0  &  0  &           0           &v^2_1&  0  &     0     &     0     &    0  \\
  0  &  0  &           0           &  0  &v^2_1&     0     &     0     &    0  \\
  0  &  0  &           0           &  0  &  0  &v^2_1+v^2_2&     0     &    0  \\
  0  &  0  &           0           &  0  &  0  &     0     &v^2_1+v^2_2&    0  \\
  0  &  0  &-\frac{v^2_2}{\sqrt{3}}&  0  &  0  &     0     &     0     &\left(4v^2_1+v^2_2\right)/3
\end{array}
 \right).
\ee
%%%%%%%%%%%%%%%%%%%%%%%%%%%%%%%%%%%
%%%%%%%%%%%%%%%%%%%%%%%%%%%%%%%%%%%

Defining the gauge boson mass eigenstates as
%%%%%%%%%%%%%%%%%%%%%%%%%%%%%%%%%%%
\be 
\left( 
\begin{array}{c}
X_\mu^{3'} \\
X_\mu^{8'}
\end{array}
  \right)
 =
\left(
\begin{array}{cc}
\cos\delta  &  \sin\delta \\
-\sin\delta &  \cos\delta
\end{array}
\right)
\left(
\begin{array}{c}
X_\mu^3 \\
X_\mu^8
\end{array}
\right),
\ee
with the mixing angle given by

\be 
\tan 2\delta = \frac{\sqrt{3} v^2_2}{2v^2_1 - v^2_2},\,\,\Longrightarrow\,\tan\delta = \frac{-2 v_1^2 + v_2^2 \pm 2 \sqrt{v_1^4 - v_1^2 v_2^2 + v_2^4}}{\sqrt{3} v_2^2},
\label{DeltaAngle}
\ee

we obtain the masses shown in Table \ref{tab:gaugemasses}. In the following, we keep only the $``+"$ solution in \eqref{DeltaAngle} corresponding to $\tan\delta$ being small and positive for $v_1^2 \gg v_2^2$.

\begin{table}[H]
\centering
\begin{tabular}{ | c | c | }
\hline 
\rowcolor[cmyk]{0.2,0.1,0.1,0.2} Gauge fields & Mass$^2$ \\ 
%\hline \hline
\hhline{|=|=|} 
\rowcolor[cmyk]{0,0,0.2,0} $X^1_\mu$ & $\frac{1}{4} g^2_X v^2_2$ \\ 
%\hline
\rowcolor[cmyk]{0,0,0.2,0} $X^2_\mu$ & $\frac{1}{4} g^2_X v^2_2$ \\ 
%\hline
\rowcolor[cmyk]{0,0.1,0.1,0.1} $X^{3'}_\mu$ & $\frac{1}{4} g^2_X v^2_2 \left( 1 - \frac{\tan\delta}{\sqrt{3}} \right)$ \\ 
%\hline
\rowcolor[cmyk]{0.1,0,0.1,0} $X^4_\mu$ & $\frac{1}{4} g^2_X v^2_1$ \\ 
%\hline
\rowcolor[cmyk]{0.1,0,0.1,0} $X^5_\mu$ & $\frac{1}{4} g^2_X v^2_1$ \\ 
%\hline
\rowcolor[cmyk]{0.2,0.0,0,0.0} $X^6_\mu$ & $\frac{1}{4} g^2_X \left( v^2_1 + v^2_2 \right) $ \\ 
%\hline
\rowcolor[cmyk]{0.2,0.0,0,0.0} $X^7_\mu$ & $\frac{1}{4} g^2_X \left( v^2_1 + v^2_2 \right) $ \\ 
%\hline
\rowcolor[cmyk]{0.1,0.0,0,0.1} $X^{8'}_\mu$ & $\frac{1}{3} g^2_X v^2_1 \left( 1 - \frac{\tan\delta}{\sqrt{3}} \right)^{-1}$ \\ 
\hline
\end{tabular}
\caption{Dark gauge boson masses.}
\label{tab:gaugemasses}
\end{table}

In addition to the above gauge boson mass terms, the scalar kinetic terms also give a scalar/gauge-boson mixing 
$$ig_XX_{\mu}^a(\partial^{\mu}\Phi_i)^{\dagger}T^a\Phi_i+H.c.\,=\,g_X\frac{v_2}{2}\left(\partial^{\mu}\phi_4\,X_{\mu}^6\,-\partial^{\mu}\phi_3X_{\mu}^7\right).$$

This leads to a redefinition of the two scalar and gauge fields involved according to
$$\begin{array}{cc}
\tilde{X}^6_{\mu}=X_{\mu}^6+\frac{2}{g_X}\frac{v_2}{v_1^2+v_2^2}\partial_{\mu}\phi_4\,\,,&\,\,\tilde{X}_{\mu}^7=X_{\mu}^7-\frac{2}{g_X}\frac{v_2}{v_1^2+v_2^2}\partial_{\mu}\phi_3 ,\\
\,&\,\\
\tilde{\phi}_3=\frac{v_1}{\sqrt{v_1^2+v_2^2}}\phi_3\,\,,&\,\,\tilde{\phi}_4=\frac{v_1}{\sqrt{v_1^2+v_2^2}}\phi_4.
\end{array}$$ 

The normalized masses for $X_6,\,X_7$ are the ones entering in Table \ref{tab:gaugemasses}, while the resulting masses of the canonical scalar fields $\tilde{\phi}_3$, $\tilde{\phi}_4$ are 

\ba 
M_{\tilde{\phi}_3}^2\,&=&\,\frac{1}{2}\left(\lambda_4+\lambda_5\right)\left(v_1^2 + v_2^2\right)\,, \label{phi3mass}\\
M_{\tilde{\phi}_4}^2\,&=&\,\frac{1}{2}\left(\lambda_4-\lambda_5\right)\left(v_1^2 + v_2^2\right) \label{phi4mass}\,.
\ea

For $v^2_1 \gg v^2_2$, the mixing angle $\delta$ is small and positive [cf. \eqref{DeltaAngle}], while $X_\mu^{1,2}$ and $X^{3'}_\mu$ are nearly degenerate in mass and also the lightest of the eight dark gauge bosons. In addition, because of their parities under $Z_2 \times Z'_2$ (cf. Table \ref{tab:parities}), they are stable and can therefore constitute DM candidates. Note, however, that $\tilde{\phi}_4$ and $X^{3'}_\mu$ have the same parities under $Z_2 \times Z'_2$. This means that the decay  process $X^{3'} \rightarrow \tilde{\phi}_4 + \rm SM$ is possible if $M_{\tilde{\phi}_4} < M_{X^{3'}}$, and in that case $\tilde{\phi}_4$ can be a DM candidate instead of $X^{3'}_\mu$. However, in the following we will study the case $M_{\tilde{\phi}_4} > M_{X^{3'}}$ and relegate this alternative scenario to future work.

%%%%%%%%%%%%%%%%%%%%%%%
%%%%%%%%%%%%%%%%%%%%%%%%%%%%%
\subsection{One-loop potential}
\label{OneLoopPot}
%%%%%%%%%%%%%%%%%%%%%%%%%%%%%
%%%%%%%%%%%%%%%%%%%%%%%%%%%%%

The one-loop potential, along the flat direction, at a renormalization scale $\mu = \Lambda$ where the tree-level potential is minimized, takes the form

\be 
V_1(\mathbf{n}\varphi)\,=\,A\,\varphi^4\,+\,B\,\varphi^4\,\ln(\varphi^2/\Lambda^2)\,,{\label{1LOOP}}
\ee
where the dimensionless coefficients $A,\,B$ are given (in the $\overline{MS}$ scheme) by 
\be 
\begin{split}
A &= \frac{1}{64\pi^2 \upsilon^4} \left[ \sum_{i=h_1,h_3,\tilde{\phi}_3,\tilde{\phi}_4} M^4_{i} \left( -\frac{3}{2} + \log \frac{M^2_{i}}{\upsilon^2} \right) +6 M^4_W \left( -\frac{5}{6} + \log \frac{M^2_W}{\upsilon^2}  \right) + 3 M^4_Z \left( -\frac{5}{6} + \log \frac{M^2_Z}{\upsilon^2} \right) \right.
  \\  & \qquad \qquad \left. + 3 \sum^8_{i=1} M^4_{X^i} \left( -\frac{5}{6} + \log \frac{M^2_{X^i}}{\upsilon^2}  \right) - 12 M^4_t \left( -1 + \log \frac{M^2_t}{\upsilon^2}  \right)  \right],
\end{split}
\ee

\be 
B = \frac{1}{64\pi^2 \upsilon^4} \left(\sum_{i=h_1,h_3,\tilde{\phi}_3,\tilde{\phi}_4} M^4_{i}  +6 M^4_W  + 3 M^4_Z 
  + 3 \sum^8_{i=1} M^4_{X^i} - 12 M^4_t  \right).
\label{Coeff-B}
\ee

Note that the model, with its present minimal field content, does not accommodate neutrino mass generation through a right-handed neutrino seesaw mechanism. Nevertheless, right-handed neutrinos can still be present and obtain their mass from a separate sector, the minimal example being a real scalar field that couples only to neutrinos. Of course, with the given symmetries of the model, if such a singlet exists, its couplings with the rest of the scalars cannot be forbidden \textit{a priori}. Nevertheless, it could be assumed that these couplings are quite small, in which case they would not affect the analysis of the rest of the model. 

Minimizing the one-loop effective potential, we obtain

\be 
V_1(\mathbf{n}\varphi)\,=\,B\,\varphi^4\,\left[\,\ln\left(\frac{\varphi^2}{v^2}\right)\,-\frac{1}{2}\,\right]\,.
\label{1loopeff}
\ee

An immediate consequence of the one-loop radiative corrections is to lift the pNGB mass to the nonzero value

\be 
M_{h_2}^2\,=\,\left.\frac{\partial^2V_1}{\partial\varphi^2}\right|_{\varphi=v}\,=\,\frac{1}{8\pi^2v^2}\left(M_{h_1}^4+M_{h_3}^4+M_{\tilde{\phi}_3}^4+M_{\tilde{\phi}_4}^4+6M_W^4+3M_Z^4+3 \sum^8_{i=1} M^4_{X^i}-12M_t^4\right).
\label{DarkonMass}
\ee

Finally, note that the one-loop corrections to the masses of $\tilde{\phi}_{3,4}$ are exactly zero, while the corrections to the masses of $h_{1,3}$ are very suppressed and can be safely ignored to a first approximation.\footnote{See~\cite{Lee2012} for a complete treatment in a relevant CSI model.} 
%%%%%%%%%%%%%%%%%%%%%%%%%%%%%%%%%
\section{Phenomenological analysis}
\label{sec:pheno}
%%%%%%%%%%%%%%%%%%%%%%%%%%%%%%%%%
In this section we study the phenomenological viability of the model. First we examine the interrelationship among the masses of the dark gauge bosons and scalars. Then, scanning over a range of values for the scalar couplings and the dark gauge coupling we find benchmark points that satisfy stability and perturbativity constraints, as well as bounds set by the first run of the LHC and measurements of the electroweak precision observables.

The Coleman-Weinberg mechanism is successfully realized if the mass of the dark scalar $M_{h_2}$ [cf. \eqref{DarkonMass}] turns out to be positive. For this to be true we must have $B>0$ [cf. \eqref{1loopeff}], or

\be 
M^4_{h_3} + M_{\tilde{\phi}_3}^4 + M_{\tilde{\phi}_4}^4 + 3 \, \sum^8_{i=1} M^4_{X_i} > \left( 317.26\GeV  \right)^4.
\ee

The scalar state $h_1$ (that we identify with the Higgs boson) has analogous couplings to the SM particles as a SM Higgs, but rescaled by the factor $\mathcal{R}_{1 1}$ from the rotation matrix \eqref{rotationmatrix},

\be 
g_{h_1 \chi \chi} = \mathcal{R}_{1 1} g^{\mathrm{SM}}_{h \chi \chi},
\ee
with $\chi \chi$ denoting a pair of SM particles. Constructing the signal strength parameter for $h_1$~\cite{Karam2015},

\be
\mu_{h_1} = \frac{\sigma \left( p p \rightarrow h_1 \right)}{\sigma^{\text{SM}} \left( p p \rightarrow h \right)} \frac{\text{BR} \left( h_1 \rightarrow \chi \chi \right)}{\text{BR}^{\text{SM}} \left( h \rightarrow \chi \chi \right)} \simeq \cos^2\alpha\cos^2\beta, 
\label{SignalStrength}
\ee
and employing the bound set by the first run of the LHC~\cite{Cheung:2014noa, Aad:2014eva, Khachatryan:2014jba, Falkowski2015}:

\be 
\mu_{h_1} > 0.81, \quad @\,95\% \,\, \mathrm{C.L.},
\ee
we can constrain the matrix element $\mathcal{R}_{1 1}$ as

\be 
\mathcal{R}_{11} = \cos\alpha\cos\beta > 0.9\,,
\label{LHCbound}
\ee
meaning that the angles $\alpha,\,\beta$ cannot be too large.

Another experimental constraint arises from the measurements of the oblique parameters $S$, $T$, and $U$. Setting $U=0$, we have~\cite{Olive2014a}

\be 
S = 0.00 \pm 0.08 , \qquad T = 0.05 \pm 0.07.
\label{Oblique}
\ee
In this model, the above parameters are given by the formulas presented in Appendix \ref{app:obliques}.

We can further constrain the model by requiring the stability of the scalar potential and the perturbativity of the couplings as they evolve with the renormalization scale. To this end, we consider the scalar couplings (except $\lambda_h$) and the gauge coupling $g_X$ and generate random values inside the intervals shown below,

\be
\lambda_1, \, \lambda_2, \, \lambda_3, \, \lambda_{h1}, \, \lambda_{h2}, \, \lambda_4, \, \lambda_5 \,\, \in \,\, \left[ 10^{-6},1 \right] , \quad g_X \,\, \in \,\,  \left[ 0,3 \right].
\ee
The scalar couplings are specified at the renormalization scale $\Lambda$ where the tree-level potential is minimized, whereas the dark gauge coupling is defined at the scale of the lightest dark gauge boson $g_X(M_{X^{3'}})$.

Then, we calculate the VEVs $v_1$, $v_2$ and the Higgs self-coupling $\lambda_h$ from the minimization conditions \eqref{mincond1}--\eqref{mincond4}. At the first stage, we keep only the points that reproduce the measured Higgs mass $M_{h_1} = 125.09 \pm 0.24 \GeV$. Subsequently, we solve numerically the two-loop RGEs (cf. Appendix \ref{app:RGEs}) and keep only the values of the couplings that remain perturbative up to the Planck scale and also satisfy the vacuum stability conditions \eqref{stabilityconditions1}--\eqref{stabilityconditions3}, as well as the bound set by LHC \eqref{LHCbound} and the constraints on the parameters $S$ and $T$ \eqref{Oblique}. We present five of these benchmark points in Table \ref{table:benchmarkpoints}.

\begin{table}
\centering
\begin{tabular}{ |c|c|c|c|c|c| }
\hline 
    &    BP1    &    BP2   &    BP3    &    BP4    &    BP5  \\ 
\hhline{|=|=|=|=|=|=|}
\rowcolor[cmyk]{0,0,0.2,0} $\lambda_1 \left( \Lambda \right)$ & $0.00008$ & $0.0112$ & $0.0014$ & $0.00017$ & $0.00015$  \\ 
%\hline
\rowcolor[cmyk]{0,0,0.2,0}$\lambda_2 \left( \Lambda \right)$ & $0.0706$ & $0.01073$ & $0.0689$ & $0.12129$ & $0.00126$  \\
%\hline 
\rowcolor[cmyk]{0,0,0.2,0}$\lambda_{h1} \left( \Lambda \right)$ & $0.00292$ & $0.0237$ & $0.00282$ & $0.0006$ & $0.0016$  \\
%\hline  
\rowcolor[cmyk]{0,0,0.2,0}$\lambda_{h2} \left( \Lambda \right)$ & $0.04116$ & $0.00323$ & $0.00031$ & $0.00109$ & $0.00344$  \\
%\hline 
\rowcolor[cmyk]{0,0,0.2,0}$\lambda_3 \left( \Lambda \right)$ & $0.00459$ & $0.0211$ & $0.0196$ & $0.00911$ & $0.00088$  \\
%\hline 
\rowcolor[cmyk]{0,0,0.2,0}$\lambda_4 \left( \Lambda \right)$ & $0.3104$ & $0.3317$ & $0.2878$ & $0.3363$ & $0.3564$  \\
%\hline 
\rowcolor[cmyk]{0,0,0.2,0}$\lambda_5 \left( \Lambda \right)$ & $0.0052$ & $0.000003$ & $0.000011$ & $0.13762$ & $0.00167$  \\
%\hline 
\rowcolor[cmyk]{0,0,0.2,0}$\lambda_h \left( \Lambda \right)$ & $0.13811$ & $0.13201$ & $0.12804$ & $0.12876$ & $0.13295$  \\
%\hline 
\rowcolor[cmyk]{0,0.2,0,0.1}$g_X$ & $1.25$ & $0.88$ & $0.81$ & $2.01$ & $0.29$  \\
%\hline 
\rowcolor[cmyk]{0.1,0,0.1,0}$v_h$ & $246.22$ & $246.22$ & $246.22$ & $246.22$ & $246.22$  \\
%\hline 
\rowcolor[cmyk]{0.1,0,0.1,0}$v_1$ & $3180.05$ & $882.78$ & $2365.61$ & $5272.32$ & $6610.41$  \\
%\hline 
\rowcolor[cmyk]{0.1,0,0.1,0}$v_2$ & $557.43$ & $869.86$ & $891.70$ & $1021.43$ & $3898.50$  \\
%\hline 
\rowcolor[cmyk]{0.2,0.0,0,0.0}$M_{h_1}$ & $125.07$ & $125.02$ & $125.17$ & $125.08$ & $125.14$  \\
%\hline 
\rowcolor[cmyk]{0.2,0.0,0,0.0}$M_{h_2}$ & $588.86$ & $97.82$ & $189.80$ & $2500.34$ & $227.22$  \\
%\hline 
\rowcolor[cmyk]{0.2,0.0,0,0.0}$M_{h_3}$ & $215.81$ & $184.42$ & $353.78$ & $512.43$ & $228.37$  \\
%\hline 
\rowcolor[cmyk]{0.2,0.0,0,0.0}$M_{\tilde{\phi}_3}$ & $1282.51$ & $504.70$ & $958.99$ & $2614.19$ & $3247.10$  \\
%\hline 
\rowcolor[cmyk]{0.2,0.0,0,0.0}$M_{\tilde{\phi}_4}$ & $1261.21$ & $504.69$ & $958.95$ & $1692.65$ & $3231.93$  \\
%\hline 
\rowcolor[cmyk]{0,0,0.2,0}$M_{X_{1,2}}$ & $349.65$ & $382.29$ & $361.14$ & $1028.25$ & $560.84$  \\
%\hline 
\rowcolor[cmyk]{0,0,0.2,0}$M_{X'_3}$ & $348.29$ & $314.41$ & $354.20$ & $1023.32$ & $531.48$ \\
%\hline 
\rowcolor[cmyk]{0,0,0.2,0}$M_{X_{4,5}}$ & $1994.73$ & $387.97$ & $958.07$ & $5307.55$ & $950.98$  \\
%\hline 
\rowcolor[cmyk]{0,0,0.2,0}$M_{X_{6,7}}$ & $2025.14$ & $544.67$ & $1023.88$ & $5406.23$ & $1104.05$  \\
%\hline 
\rowcolor[cmyk]{0,0,0.2,0}$M_{X'_8}$ & $2312.35$ & $544.70$ & $1127.97$ & $6158.13$ & $1158.77$  \\
%\hline 
\rowcolor[cmyk]{0.1,0,0.1,0}$\Lambda$ & $1747.67$ & $407.03$ & $834.25$ & $4704.95$ & $1838.82$  \\
\rowcolor[cmyk]{0.5,0.0,0,0.0}$\Omega_X h^2$ & $0.0365$ & $0.0670$ & $0.1136$ & $0.0952$ & $6.19$  \\
\rowcolor[cmyk]{0.5,0.0,0,0.0}$\sigma^{\rm eff}_{1,2}$ & $2.2 \times 10^{-45}$ & $1.0 \times 10^{-47}$ & $1.5 \times 10^{-47}$ & $8.7 \times 10^{-48}$ & $0$  \\
\rowcolor[cmyk]{0.5,0.0,0,0.0}$\sigma^{\rm eff}_{3}$ & $1.2 \times 10^{-44}$ & $7.7 \times 10^{-46}$ & $2.8 \times 10^{-46}$ & $5.5 \times 10^{-47}$ & $1.5 \times 10^{-46}$  \\
\hline 
\end{tabular}
\caption{Benchmark points for the model parameters that satisfy the stability and perturbativity constraints, as well as the bounds set by LHC and measurements of the oblique parameters. The VEVs, the masses, and $\Lambda$ are in $\rm GeV$ units. For completeness, we have also included the values of the total relic density of $X^{1,2,3'}$ and their effective scattering cross sections off a nucleon (in $\rm cm^2$ units) which we discuss in Secs. \ref{subsec:RD} and \ref{subsec:DD}.}
\label{table:benchmarkpoints}
\end{table}

Most of these benchmark points (BPs) contain values for the dark VEVs for which $v_1^2 \gg v_2^2$. This results in the masses of the dark gauge bosons $X^1_\mu$, $X^2_\mu$, $X^{3'}_\mu$ being nearly degenerate, while the masses of the rest of the dark gauge bosons are well above them. Nonetheless, in BP2, we have also included the case $v_1 \simeq v_2$. In this case, the mass of $X^{3'}_\mu$ is fairly lower than the masses of $X^1_\mu$ and $X^2_\mu$, which are now close to the masses of $X^4_\mu$ and $X^5_\mu$, while the masses of $X^6_\mu$ and $X^7_\mu$ become nearly degenerate with the mass of $X^{8'}_\mu$. Therefore, in the case $v_1 \simeq v_2$, we have

\be 
M^2_{X^{3'}} \simeq \frac{2}{3} M^2_{X^{1,2}} \simeq \frac{2}{3} M^2_{X^{4,5}} \simeq \frac{1}{3} M^2_{X^{6,7}} \simeq \frac{1}{3} M^2_{X^{8'}}.
\label{eq:secondlimitmasses}
\ee
As we will see in the next section, the case $v_1 \simeq v_2$ is distinct in its dark matter analysis.

Regarding the scalar bosons and the pNGB $h_2$ in particular, we observe that its mass depends highly on the values of the VEVs $v_1$, $v_2$ and the dark gauge coupling $g_X$, or equivalently on the masses of the dark gauge bosons and the rest of the scalars [cf. \eqref{DarkonMass}]. For example, large values for the VEVs and $g_X$ produce a large mass for $h_2$, as can be seen from BP4 in Table \ref{table:benchmarkpoints}.

\begin{figure}
\includegraphics[width=.5\textwidth]{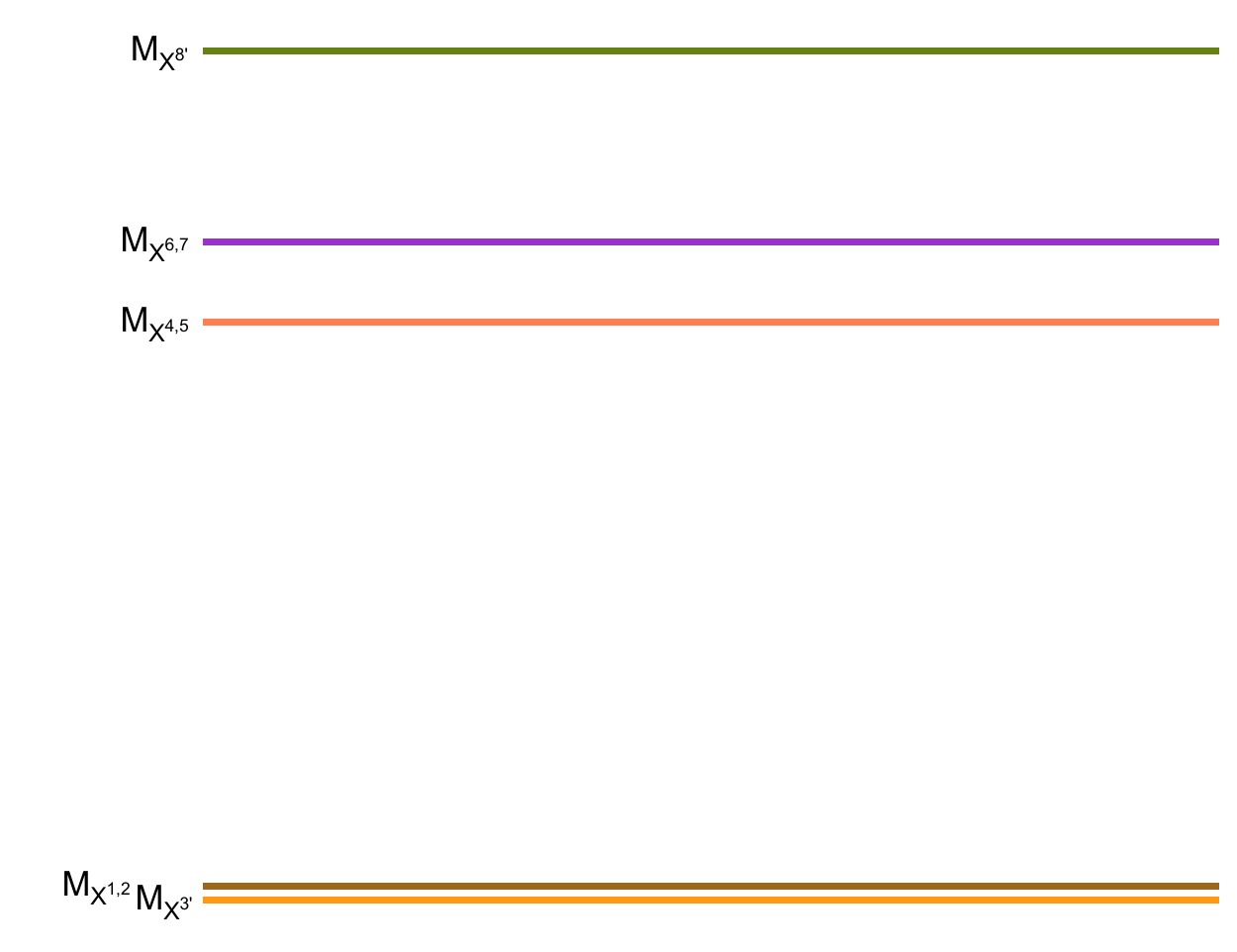}
\includegraphics[width=.5\textwidth]{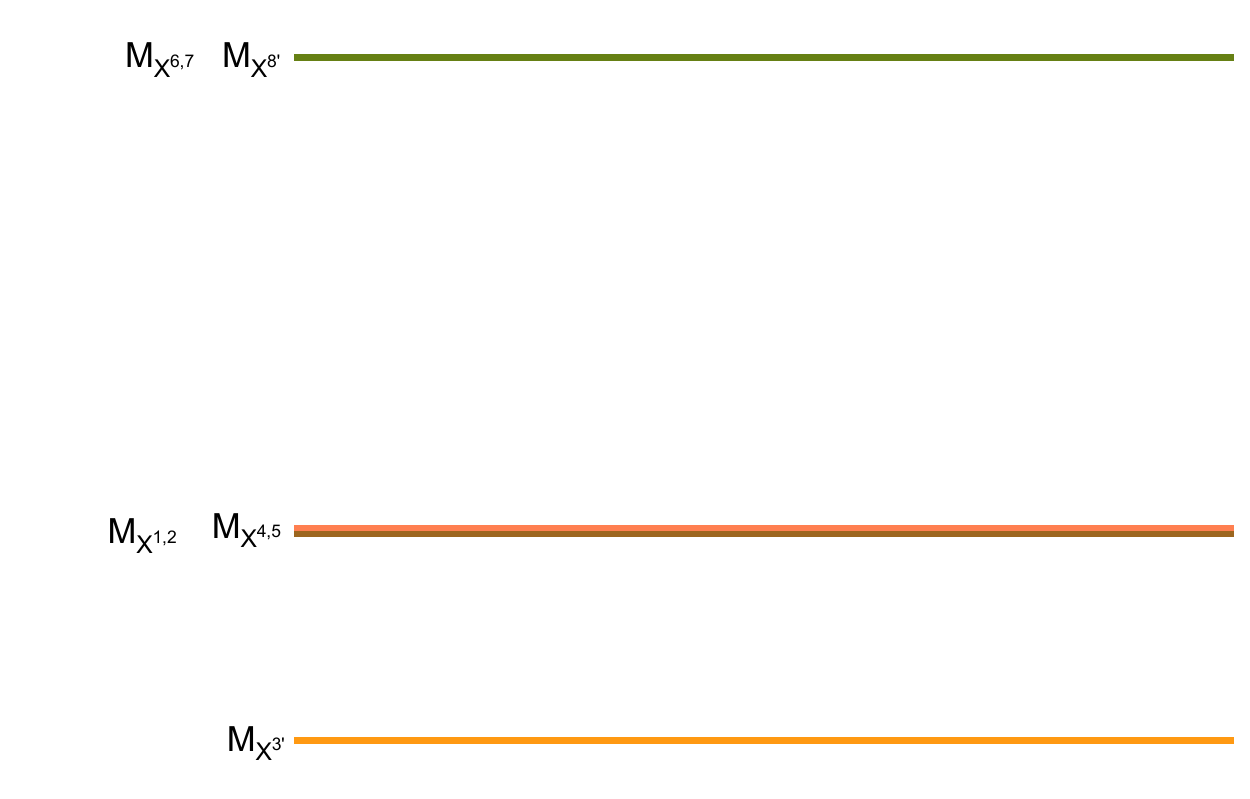}
\caption{Mass spectra of dark gauge bosons for the cases $v_1^2 \gg v_2^2$ (left) and $v_1 \simeq v_2$ (right).}
\label{fig:massspectra}
\end{figure}

Finally, the dark gauge boson mass spectrum for both cases $v_1^2 \gg v_2^2$ and $v_1 \simeq v_2$ is shown schematically in Fig. \ref{fig:massspectra}.

%%%%%%%%%%%%%%%%%%%%%%%
%%%%%%%%%%%%%%%%%%%%%%%%%%%%%%%%%
\section{Dark matter analysis}
\label{sec:DM}
%%%%%%%%%%%%%%%%%%%%%%%%%%%
%%%%%%%%%%%%%%%%%%%%%%%%%%%
Recent astrophysical measurements~\cite{Ade2015} have corroborated the now well-established fact that $\sim 80 \%$ of the nonrelativistic matter in the Universe is in a form that remains a mystery to us and cannot be explained by the known particles and forces. This ``dark matter" (DM) could be constituted of scalar bosons, fermions, vector bosons, a combination of the above, or even something more exotic. Here we will focus on vector DM~\cite{Hambye2009, Hambye2010, Zhang2010, Arina2010, Bhattacharya2012, Chiang2014, Boehm2014, Baek2013a, DEramo2013, Farzan2012, Djouadi2012b, Choi2013a, Baek2014a, Baek2014, Ko2014, Balazs2014, Farzan2014, Fraser2015, Chen2015c, Chao2015, Duch2015a, DiChiara2015b, Gross2015, Bernal2016, Sage2016, Khoze2014b, Bian2015, Chen2015, Chen2015a, Chen2015b, Hambye2013, Carone2013, Khoze2014, Pelaggi2015, Karam2015, DiFranzo2016, Ko2016, Karam:2016bhq}.

Whatever the case may be, a DM candidate particle should be stabilized by some kind of symmetry, such that it may not decay to the SM particles. The simplest possibility of a stabilizing symmetry is that of a $Z_2$ discrete symmetry. A neutral and weakly interacting massive particle can be a DM candidate if it is the lightest $Z_2$-odd particle in a given model. In order to accommodate more DM candidates, one should consider a $Z_N$ ($N\geq 4$) or a product of two or more $Z_2$'s as the stabilizing symmetry.

The intrinsic $Z_2\times Z_2'$ symmetry of the dark sector of the model, not shared by the SM fields, singles out the particles with nontrivial signatures under this symmetry as a stable sector without any other symmetry requirements. The lightest of the dark gauge bosons then, are possible dark matter candidates. Under our assumptions, the lightest of them are the dark gauge bosons $X^1_\mu,\,X^2_\mu$ and $X^{3'}_\mu$.

The present model allows for various processes that are able to change the number density of dark matter particles. These are the following:

\begin{enumerate}[(a)]
\item \textit{Annihilation into SM.} All dark gauge bosons interact with the scalars $h_i$ ($i=1,2,3$), which in turn communicate with the SM fields. Thus, the DM candidates $X_\mu^{1,2,3'}$ can annihilate to the SM particles through the Higgs portal.
\item \textit{Semiannihilation.} The non-Abelian nature of the extra gauge symmetry allows the processes $X^a X^b \rightarrow X^c h_i$ to occur. In this case, the final number of DM particles is one less than the initial number, as opposed to the case of annihilations where the DM number of particles is changed by two units. Semiannihilation processes are of great interest regarding DM phenomenology since they can dominate in much of the parameter space.
\item \textit{Coannihilation.} This kind of process has been thoroughly investigated in the context of supersymmetric DM models.\footnote{See, for example, Ref.~\cite{Ellis2016} and references therein.} There, the lightest neutralino particle (LSP) is a DM candidate and can potentially coannihilate with the next-to-lightest supersymmetric particle (NLSP) if their respective masses are close enough. A similar situation arises in the dark sector of the model under consideration when $v_1 \simeq v_2$, since in that case the masses of the DM candidates $X_\mu^1$ and $X_\mu^2$ are close to those of $X_\mu^4$ and $X_\mu^5$ (cf. Fig. \ref{fig:massspectra}) and may in principle coannihilate with them through the processes $X^{1} X^{4,5} \rightarrow X^{7,6} h_i$ and $X^{2} X^{4,5} \rightarrow X^{6,7} h_i$. Notice, however, that we cannot employ the usual condition between the LSP and NLSP(s) number densities before, during, and after freeze-out, namely $n_i/n_j\,=\,n_i^{eq}/n_j^{eq}$, since its validity cannot be guaranteed when semiannihilations are also involved (see Ref.~\cite{DEramo2010} for more details).
\item \textit{DM conversion.} In multicomponent DM systems the various DM candidates have different masses in general. Then, if the relevant interactions are allowed, a DM species may be converted to another. In this model the three DM candidates $X_\mu^{1,2,3'}$ are nearly degenerate in mass, and such processes $\left( X^{1,2} X^{1,2} \rightarrow X^{3'} X^{3'} \right)$ are generally phase space suppressed. However, again in the limiting case $v_1 \simeq v_2$ the mass splitting of $X_\mu^{3'}$ with regard to $X_\mu^1$ and $X_\mu^2$ can have a significant effect in today's number density of these DM species.
\end{enumerate}
%%%%%%%%%%%%%%%%%%%%%%%
\subsection{Boltzmann equations and relic density}
\label{subsec:RD}
%%%%%%%%%%%%%%%%%%%%%%%
In order to determine the present day abundance of the DM species we need to solve a coupled set of Boltzmann equations involving the number densities of the dark sector particles. These equations can be written in a compact form as

\be
\frac{dn_a}{dt}\,+\,3H\,n_a\,=\,{\cal{C}}_a\,\,\,\,\,\,\,\,\,(a=1,2,3')\,,{\label{BOLTZ-0}}
\ee
with $H$ being the Hubble parameter and ${\cal{C}}_a\,=\,\sum_{bcd}{\cal{C}}_{a b\rightarrow cd}$ being the collision rate of all possible $2 \rightarrow 2$ processes for a given species that can change its number density.
We can relate the collision rate of a reaction with its inverse by making use of the detailed balance equation

\be 
{\mathcal{C}}_{ab\rightarrow cd}\,=\,-\langle\sigma_{ab\rightarrow cd}v_r\rangle\left(n_an_b-n_cn_d\frac{\bar{n}_a\bar{n}_b}{\bar{n}_c\bar{n}_d}\right)\,=\,+\langle\sigma_{cd\rightarrow ab}v_r\rangle\left(n_cn_d-n_an_b\frac{\bar{n}_c\bar{n}_d}{\bar{n}_a\bar{n}_b}\right)\,, 
\ee
where $\bar{n} \equiv n^{\mathrm{eq}}$ is the equilibrium number density and $\langle\sigma_{ab\rightarrow cd}v_r\rangle$ is the thermally averaged cross section times the relative velocity of the DM particles. It is given by the general formula~\cite{Gondolo1991a, Edsjo1997, Ellis2000} 

\be 
\langle \sigma_{ab\rightarrow cd} v_r \rangle = \frac{1}{2 m^2_a m^2_b T K_2(m_a/T)K_2(m_b/T)} \int^\infty_{(m_a + m_b)^2} ds K_1 (\sqrt{s}/T) p_{\mathrm{in}}(s) w(s) ,
\label{THERM-1}
\ee
where $w(s)\,=\,E_aE_b\sigma_{ab\rightarrow cd}v_r$. The cross section for a given process $a + b \rightarrow c + d$ is

\be 
\sigma_{ab\rightarrow cd}v_r\,=\,\frac{1}{1 + \delta_{cd}}\frac{p_{\mathrm{out}}(s)}{32\pi s p_{\mathrm{in}}(s)}\int\,d\cos\theta\,|{\mathcal{M}}_{ab\rightarrow cd}|^2\,,{\label{CROSS}}
\ee
with $|{\mathcal{M}}|^2$ denoting the spin summed and polarization averaged matrix element squared. In Eq. \eqref{THERM-1}, $K_\nu(z)$ stands for the modified Bessel functions. The  general expressions for the kinematical variables contained in \eqref{THERM-1} and \eqref{CROSS} are provided in Appendix \ref{app:kin}.

We may now proceed to obtain the relic abundance of the DM candidates by solving numerically the set of Boltzmann equations. In order to write down the system of coupled equations, we need to identify the reactions which modify the number of $X_\mu^1$, $X_\mu^2$, and $X_\mu^{3'}$ particles. Since $M_{X_1} = M_{X_2} > M_{X_{3'}}$, the number densities satisfy $n_1 = n_2 \neq n_3$. It should also be clear that $\langle \sigma v_r \rangle_{11\rightarrow \chi\chi'} = \langle \sigma v_r \rangle_{22\rightarrow \chi\chi'} \neq \langle \sigma v_r \rangle_{33\rightarrow \chi\chi'}$, $\langle \sigma v_r \rangle_{12\rightarrow 3\chi} = \langle \sigma v_r \rangle_{21\rightarrow 3\chi} \neq \langle \sigma v_r \rangle_{13\rightarrow 2\chi} = \langle \sigma v_r \rangle_{31\rightarrow 2\chi} = \langle \sigma v_r \rangle_{23\rightarrow 1\chi} = \langle \sigma v_r \rangle_{32\rightarrow 1\chi}$, and $\langle \sigma v_r \rangle_{11\rightarrow 33} = \langle \sigma v_r \rangle_{22\rightarrow 33} \neq \langle \sigma v_r \rangle_{33\rightarrow 11} = \langle \sigma v_r \rangle_{33\rightarrow 22}$, where, for example, $\langle \sigma v_r \rangle_{12\rightarrow 3\chi}$ is short for $\langle \sigma v_r \rangle_{X_1 X_2\rightarrow X_{3'}\chi}$, etc., and $\chi \chi'$ denotes $\rm SM \, \rm SM$ and $h_i h_j$ pairs when these are kinematically allowed. 

The processes which modify the number of $X_\mu^{1,2}$ particles are

\be
X^{1,2} X^{1,2} \rightarrow \chi \chi' , \qquad X^{1,2} X^{2,1} \rightarrow X^{3',8'} h_i , \qquad X^{1,2} X^{3'} \rightarrow X^{2,1} h_i , \qquad X^{1,2} X^{1,2} \rightarrow X^{3'} X^{3'} ,
\ee
whereas the ones which modify the number of $X^{3'}$ particles are

\be
X^{3'} X^{3'} \rightarrow \chi \chi' , \qquad X^{1,2} X^{3'} \rightarrow X^{2,1} h_i , \qquad X^{3'} h_i \rightarrow X^{1,2} X^{2,1} , \qquad X^{3'} X^{3'} \rightarrow X^{1,2} X^{1,2} .
\ee

The collision operators for the processes which modify the number of $X_\mu^1$ and $X_\mu^2$ particles are

\be 
\begin{split}
\mathcal{C}_{1 1 \rightarrow \chi\chi'} = & - \langle \sigma v_r\rangle_{1 1 \rightarrow \chi\chi'} \left[n_1^2 - \overline{n}^2_1\right] = \mathcal{C}_{2 2 \rightarrow \chi\chi'}, \\ 
\mathcal{C}_{1 2 \rightarrow 3 h_i} = & - \langle \sigma v_r\rangle_{1 2 \rightarrow 3 h_i} \left[n_1 n_2 - \overline{n}_1 \overline{n}_2 \frac{n_3}{\overline{n}_3}\right] = \mathcal{C}_{2 1 \rightarrow 3 h_i} , \\
\mathcal{C}_{1 2 \rightarrow 8 h_i} = & - \langle \sigma v_r\rangle_{1 2 \rightarrow 8 h_i} \left[n_1 n_2 - \overline{n}_1 \overline{n}_2 \right] = \mathcal{C}_{2 1 \rightarrow 8 h_i} , \\
\mathcal{C}_{1 3 \rightarrow 2 h_i} = & - \langle \sigma v_r\rangle_{1 3 \rightarrow 2 h_i} \left[n_1 n_3 - \overline{n}_1 \overline{n}_3 \frac{n_2}{\overline{n}_2}\right] = \mathcal{C}_{2 3 \rightarrow 1 h_i} , \\
\mathcal{C}_{1 1 \rightarrow 3 3} = & - \langle \sigma v_r\rangle_{1 1 \rightarrow 3 3} \left[n_1^2 - n^2_3 \frac{\overline{n}^2_1}{\overline{n}^2_3}\right] = \mathcal{C}_{2 2 \rightarrow 3 3}, \\
\mathcal{C}_{1 h_i \rightarrow 2 3} = & + \langle \sigma v_r\rangle_{2 3 \rightarrow 1 h_i} \left[n_2 n_3 - \overline{n}_2 \overline{n}_3 \frac{n_1}{\overline{n}_1}\right] = \mathcal{C}_{2 h_i \rightarrow 1 3} ,
\end{split}
\ee
whereas the ones which modify the number of $X_\mu^{3'}$ particles are

\be 
\begin{split}
\mathcal{C}_{3 3 \rightarrow \chi\chi'} = & - \langle \sigma v_r\rangle_{3 3 \rightarrow \chi\chi'} \left[n_3^2 - \overline{n}^2_3\right] , \\ 
\mathcal{C}_{1 3 \rightarrow 2 h_i} = & - \langle \sigma v_r\rangle_{1 3 \rightarrow 2 h_i} \left[n_1 n_3 - \overline{n}_1 \overline{n}_3 \frac{n_2}{\overline{n}_2}\right] = \mathcal{C}_{2 3 \rightarrow 1 h_i} , \\
\mathcal{C}_{3 3 \rightarrow 1 1} = & + \langle \sigma v_r\rangle_{1 1 \rightarrow 3 3} \left[n_1^2 - n^2_3 \frac{\overline{n}^2_1}{\overline{n}^2_3}\right] = \mathcal{C}_{3 3 \rightarrow 2 2} , \\
\mathcal{C}_{3 h_i \rightarrow 1 2} = & + \langle \sigma v_r\rangle_{1 2 \rightarrow 3 h_i} \left[n_1 n_2 - \overline{n}_1 \overline{n}_2 \frac{n_3}{\overline{n}_3}\right] .
\end{split}
\ee

As discussed above, in the case $v_1 \simeq v_2$, the particles $X_\mu^{4,5}$ are thermally available to $X_\mu^{1,2}$ and may coannihilate with them. We therefore also have to include them in our analysis. The collision operators for the processes which change the number of $X_\mu^{4,5}$ particles are\footnote{Of course, these reactions also change the number of $X_\mu^{1,2,3'}$ particles. Also, we have assumed that the heavier dark gauge bosons $X_\mu^{6,7,8'}$ have already decayed to the lighter ones.}

\be 
\begin{split}
\mathcal{C}_{4 4 \rightarrow \chi\chi'} = & - \langle \sigma v_r\rangle_{4 4 \rightarrow \chi\chi'} \left[n_4^2 - \overline{n}^2_4\right] = \mathcal{C}_{5 5 \rightarrow \chi\chi'}, \\ 
\mathcal{C}_{1 4 \rightarrow 7 h_i} = & - \langle \sigma v_r\rangle_{1 4 \rightarrow 7 h_i} \left[n_1 n_4 - \overline{n}_1 \overline{n}_4 \right] = \mathcal{C}_{1 5 \rightarrow 6 h_i} = \mathcal{C}_{2 4 \rightarrow 6 h_i} = \mathcal{C}_{2 5 \rightarrow 7 h_i} , \\
\mathcal{C}_{4 4 \rightarrow 1 1} = & - \langle \sigma v_r\rangle_{4 4 \rightarrow 1 1} \left[n_4^2 - n^2_1 \frac{\overline{n}^2_4}{\overline{n}^2_1}\right] = \mathcal{C}_{4 4 \rightarrow 2 2}= \mathcal{C}_{5 5 \rightarrow 1 1}= \mathcal{C}_{5 5 \rightarrow 2 2}, \\
\mathcal{C}_{4 4 \rightarrow 3 3} = & - \langle \sigma v_r\rangle_{4 4 \rightarrow 3 3} \left[n_4^2 - n^2_3 \frac{\overline{n}^2_4}{\overline{n}^2_3}\right] = \mathcal{C}_{5 5 \rightarrow 3 3}
\end{split}
\ee

Next, let us define 

\be 
Y_a \equiv \frac{n_a}{s}, \qquad x \equiv \frac{M_{X^{3'}}}{T}, \qquad \mathcal{Z}_{ab \rightarrow cd}(x)  \equiv \frac{s(x=1)}{H(x=1)} \langle \sigma v_r\rangle_{ab \rightarrow cd}\,,
\ee
where $ H = \sqrt{\frac{4 \pi^3 g_\star}{45}} \, \frac{T^2}{M_{\mathrm{Pl}}}$, $g_\star \simeq g_{\star s}$ is the number of effective relativistic degrees of freedom, and $s = \frac{2 \pi^2 g_{\star s}}{45} \, T^3$ is the entropy density. Then, we may finally write down the coupled set of Boltzmann equations in dimensionless variables as

\be 
\begin{split}
\frac{d Y_1}{d x} &= - \frac{1}{x^2} \left\{ \mathcal{Z}_{1 1 \rightarrow \chi\chi'} \left[Y_1^2 - \overline{Y}^2_1\right] + \mathcal{Z}_{1 2 \rightarrow 3 h_i} 
\left[Y_1 Y_2 - \overline{Y}_1 \overline{Y}_2 \frac{Y_3}{\overline{Y}_3}\right] \right. \\ & \left. \qquad \qquad +~\mathcal{Z}_{1 2 \rightarrow 8 h_i} 
\left[Y_1 Y_2 - \overline{Y}_1 \overline{Y}_2 \right] + \mathcal{Z}_{1 3 \rightarrow 2 h_i} 
\left[Y_1 Y_3 - \overline{Y}_1 \overline{Y}_3 \frac{Y_2}{\overline{Y}_2}\right] \right. \\ & \left. \qquad \qquad -~ \mathcal{Z}_{2 3 \rightarrow 1 h_i} \left[Y_2 Y_3 - \overline{Y}_2 \overline{Y}_3 \frac{Y_1}{\overline{Y}_1}\right] + \mathcal{Z}_{1 1 \rightarrow 3 3} \left[Y_1^2 - Y^2_3 \frac{\overline{Y}^2_1}{\overline{Y}^2_3}\right] \right. \\ & \left. 
\qquad \qquad -~ \mathcal{Z}_{4 4 \rightarrow 1 1} \left[Y_4^2 - Y^2_1 \frac{\overline{Y}^2_4}{\overline{Y}^2_1}\right] - \mathcal{Z}_{5 5 \rightarrow 1 1} \left[Y_5^2 - Y^2_1 \frac{\overline{Y}^2_5}{\overline{Y}^2_1}\right] \right. \\ & \left. 
\qquad \qquad +~ \mathcal{Z}_{1 4 \rightarrow 7 h_i} 
\left[Y_1 Y_4 - \overline{Y}_1 \overline{Y}_4 \right] + \mathcal{Z}_{1 5 \rightarrow 6 h_i} 
\left[Y_1 Y_5 - \overline{Y}_1 \overline{Y}_5 \right] \right\},
\end{split}
\label{eq:BEsystem1}
\ee

\be 
\frac{d Y_2}{d x} = \frac{d Y_1}{d x}(1\leftrightarrow 2,4\leftrightarrow 5,7\leftrightarrow 6) \label{eq:BEsystem2}
\ee

\be
\begin{split}
\frac{d Y_3}{d x} &= - \frac{1}{x^2} \left\{ \mathcal{Z}_{3 3 \rightarrow \chi\chi'} \left[Y_3^2 - \overline{Y}^2_3\right] + \mathcal{Z}_{1 3 \rightarrow 2 h_i} 
\left[Y_1 Y_3 - \overline{Y}_1 \overline{Y}_3 \frac{Y_2}{\overline{Y}_2}\right] \right. \\ & \left. \qquad \qquad +~\mathcal{Z}_{2 3 \rightarrow 1 h_i} 
\left[Y_2 Y_3 - \overline{Y}_2 \overline{Y}_3 \frac{Y_1}{\overline{Y}_1}\right] - \mathcal{Z}_{1 1 \rightarrow 3 3} \left[Y_1^2 - Y^2_3 \frac{\overline{Y}^2_1}{\overline{Y}^2_3}\right] \right. \\ & \left. \qquad \qquad -~ \mathcal{Z}_{2 2 \rightarrow 3 3} \left[Y_2^2 - Y^2_3 \frac{\overline{Y}^2_2}{\overline{Y}^2_3}\right] - \mathcal{Z}_{4 4 \rightarrow 3 3} \left[Y_4^2 - Y^2_3 \frac{\overline{Y}^2_4}{\overline{Y}^2_3}\right] \right. \\ & \left. 
\qquad \qquad -~ \mathcal{Z}_{5 5 \rightarrow 3 3} \left[Y_5^2 - Y^2_3 \frac{\overline{Y}^2_5}{\overline{Y}^2_3}\right] - 2 \, \mathcal{Z}_{1 2 \rightarrow 3 h_i} \left[Y_1 Y_2 - \overline{Y}_1 \overline{Y}_2 \frac{Y_3}{\overline{Y}_3}\right] \right\}.
\end{split}
\label{eq:BEsystem3}
\ee

\be 
\begin{split}
\frac{d Y_4}{d x} &= - \frac{1}{x^2} \left\{ \mathcal{Z}_{4 4 \rightarrow \chi\chi'} \left[Y_4^2 - \overline{Y}^2_4\right] + \mathcal{Z}_{4 4 \rightarrow 1 1} \left[Y_4^2 - Y^2_1 \frac{\overline{Y}^2_4}{\overline{Y}^2_1}\right] \right. \\ & \left. \qquad \qquad +~\mathcal{Z}_{4 4 \rightarrow 2 2} \left[Y_4^2 - Y^2_2 \frac{\overline{Y}^2_4}{\overline{Y}^2_2}\right] + \mathcal{Z}_{4 4 \rightarrow 3 3} \left[Y_4^2 - Y^2_3 \frac{\overline{Y}^2_4}{\overline{Y}^2_3}\right] \right. \\ & \left. \qquad \qquad +~ \mathcal{Z}_{1 4 \rightarrow 7 h_i} 
\left[Y_1 Y_4 - \overline{Y}_1 \overline{Y}_4 \right] + \mathcal{Z}_{2 4 \rightarrow 6 h_i} 
\left[Y_2 Y_4 - \overline{Y}_2 \overline{Y}_4 \right] \right\},
\end{split}
\label{eq:BEsystem4}
\ee

\be 
\frac{d Y_5}{d x} = \frac{d Y_4}{d x}(4\leftrightarrow 5,1\leftrightarrow 2,7\leftrightarrow 6). \label{eq:BEsystem5}
\ee

The equilibrium yields $\overline{Y}_a \equiv \frac{\overline{n}_a}{s}$ are given by
\ba 
\overline{Y}_3 &=& \frac{\hat{g}_X}{g_{\star s}} \frac{45}{4\pi^4} x^2 K_2(x), \\
\overline{Y}_{1,2} &=& \frac{\hat{g}_X}{g_{\star s}} \frac{45}{4\pi^4} r^2_{1,2} x^2 K_2( r_{1,2} x), \\
\overline{Y}_{4,5} &=& \frac{\hat{g}_X}{g_{\star s}} \frac{45}{4\pi^4} r^2_{4,5} x^2 K_2( r_{4,5} x),
\ea
where we have defined $r_{1,2}\equiv \frac{M_{X^{1,2}}}{M_{X^{3'}}}$, $r_{4,5}\equiv \frac{M_{X^{4,5}}}{M_{X^{3'}}}$ and $\hat{g}_X=3$ are the spin degrees of freedom of the dark gauge bosons.
We have numerically solved this system using \texttt{Mathematica} and we have also employed the packages \texttt{FeynArts/FormCalc}~\cite{Hahn2001, Hahn1999} in order to produce analytic results for the various cross sections involved. Finally, we have obtained the total relic density of the $X_\mu^{1,2,3'}$ particles
\be 
\Omega_{X} h^2 = \Omega_{X^1} h^2 + \Omega_{X^2} h^2 + \Omega_{X^{3'}} h^2 ,
\label{eq:RelicDensity}
\ee
where 
\be 
\Omega_{X^a} h^2 = \frac{M_{X^a} s_0 Y^a(\infty)}{\rho_c / h^2} ,
\ee
with $s_0 = 2890 \,\, \rm cm^{-3}$ and $\rho_c / h^2 = 1.05 \times 10^{-5}\GeV / \rm cm^3$.
Equation \eqref{eq:RelicDensity} has to be compared with the measured DM relic density $\Omega_{\mathrm{DM}} h^2 = 0.1197 \pm 0.0022$~\cite{Ade2015}. Next, we further explore the cases $v_1^2\gg v_2^2$ and $v_1 \simeq v_2$.
%%%%%%%%%%%%%%%%%%%%%%%%%%%%%%%%%%%%
\subsubsection{Case $v_1^2\gg v_2^2$}
%%%%%%%%%%%%%%%%%%%%%%%%%%%%%%%%%%%%
In this case, as stated above, the masses of the DM candidates $X^1$, $X^2$, and $X^{3'}$ are nearly degenerate, while the masses of $X^4$ and $X^5$ are well above those of $X^1$ and $X^2$. Therefore, coannihilation effects play no significant role in the final relic density of $X^{1,2,3'}$. However, even though the mass splitting between $M_{X^1}=M_{X^2}$ and $M_{X^{3'}}$ is small, the DM conversion processes $X^{1,2} X^{1,2} \rightarrow X^{3'} X^{3'}$ can lower the number density of $X^{1}$ and $X^2$ and enhance that of $X^{3'}$, rendering $X^{3'}$ the predominant DM component. 

To get a feeling of the effect of DM conversion, we set the parameters of the model according to BP1 of Table \ref{table:benchmarkpoints} and solve numerically the Boltzmann equations \eqref{eq:BEsystem1}--\eqref{eq:BEsystem3} (omitting the coannihilation terms), thus obtaining the solutions for the yields $Y_{1,2}$ and $Y_3$ with respect to $x = M_{X^{3'}}/T$. 

\begin{figure}
\includegraphics[width=.5\textwidth]{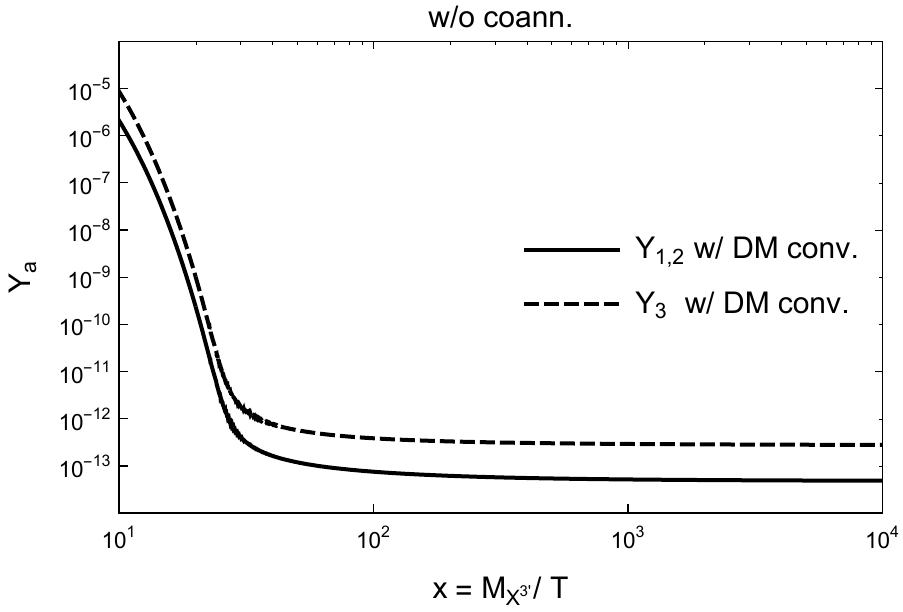}
\includegraphics[width=.5\textwidth]{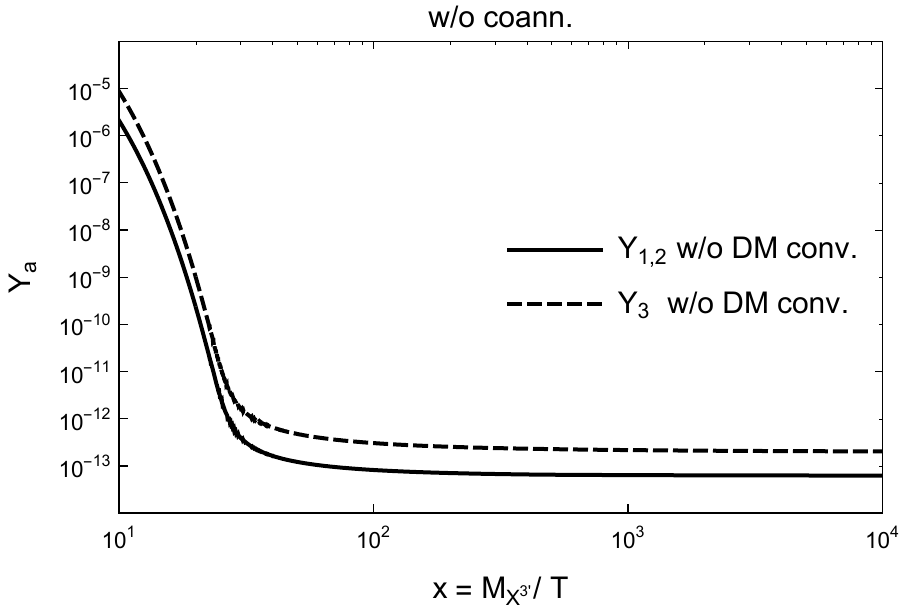}
\caption{The yields $Y_{1,2}$ and $Y_3$ in terms of $x = M_{X^{3'}}/T$ for BP1. The right plot has been obtained neglecting the DM conversion terms in the Boltzmann equations. These terms are included in the left plot. The DM conversion process reduces the final number density of the $X^1$ and $X^2$ particles since some of them are converted to $X^{3'}$.}
\label{fig:YieldPlotsBP1}
\end{figure}

In Fig. \ref{fig:YieldPlotsBP1} we plot these solutions with the DM conversion processes switched on (left) and switched off (right). When the DM conversion is switched off, the final yields are closer together, with the separation attributed to the slightly different masses between $X^{1,2}$ and $X^{3'}$, as well as to the mixing between $X^{3'}-X^{8'}$ which results in more Feynman diagrams contributing to the annihilation processes $X^{3'} X^{3'} \rightarrow h_i h_j$ and the semiannihilation processes $X^{1,2} X^{2,1} \rightarrow X^{3'} h_i$.\footnote{In the non-CSI version of this model considered in Ref. \cite{Gross2015}, the authors performed their $SU(3)$ DM analysis under the simplified assumption that the rest of the dark sector particles do not contribute to DM annihilation. They also only included the couplings of the DM candidates to the two lightest scalar bosons $h_1$, $h_2$ and not the heavier $\mathcal{H}$ in their notation. Here, we include all possible couplings and Feynman diagrams relevant to the relic densities of $X^{1,2,3'}$.} On the other hand, the separation of the final yields is larger when the DM conversion processes are switched on, since more $X^1$ and $X^2$ particles have annihilated and have been converted to $X^{3'}$; a reaction that continues to occur to some extent even after freeze-out. In the case without DM conversion, the particles $X^1$, $X^2$, and $X^{3'}$ comprise $19 \%$, $19 \%$, and $62 \%$ of the total relic density respectively, while in the case with DM conversion they comprise $13 \%$, $13 \%$, and $74 \%$ of the total relic density, respectively.

\begin{figure}
\centering
\includegraphics[width=0.68\textwidth]{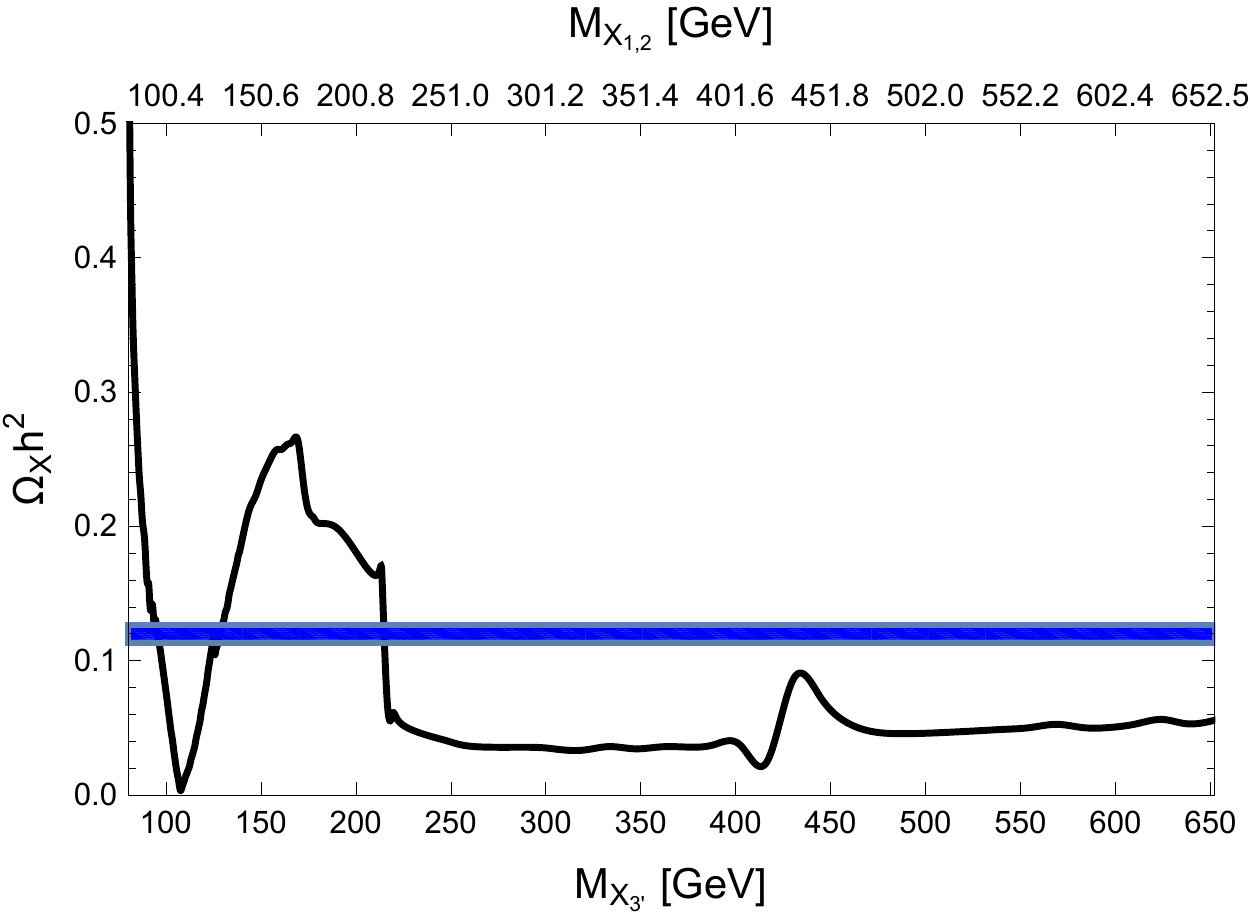}
\caption{The total relic density of $X^{1,2}$ and $X^{3'}$ as a function of the dark gauge coupling $g_X$ for BP1. The blue band corresponds to the observed DM relic density within $3\sigma$.}
\label{fig:RDBP1}
\end{figure}

In Fig. \ref{fig:RDBP1} we fix again the model parameters as in BP1, but this time we leave the extra gauge coupling $g_X$ free and scan over it, ergo obtaining the total relic density $\Omega_X h^2$ of the DM candidates. We first observe a resonant dip around $110\GeV$ which corresponds to $M_{X^{3'}} \simeq M_{X^{1,2}} = M_{h_3} / 2$. Then the relic density grows until $\sim 175\GeV$ where the $t \bar{t}$ channel opens up. After that, there is a steep decrease around $M_{h_3} \simeq 215\GeV$ where all the annihilation channels $X^a X^a \rightarrow h_3 h_3$ and the semiannihilation channels $X^a X^b \rightarrow X^c h_3$ become kinematically available. This point crosses the observed DM relic density (blue band in Fig. \ref{fig:RDBP1}) and corresponds to $g_X = 0.78$ (which also satisfies the constraints discussed in Sec. \ref{sec:pheno}). 

Above $M_{h_3}$, one may expect that the relic density would decrease monotonically. This can be understood as follows: every vertex containing three dark gauge boson legs is proportional to $g_X$ while every vertex containing two dark gauge bosons and one or two scalar bosons is proportional to $g_X^2$. Therefore, $\left< \sigma v_r \right> \propto g^2_X $, or $\Omega_X h^2 \propto 1/g^2_X$. This indicates that the relic density should decrease as we increase $g_X$ (and therefore $M_{X^{1,2,3'}}$). Nevertheless, the mass of the pNGB $M_{h_2}$ depends on all the masses of the model [cf. \eqref{DarkonMass}]. This means that as $g_X$ grows, so do the dark gauge boson masses and consequently $M_{h_2}$. This effect tends to counterbalance the expected decrease of $\Omega_X h^2$. On the other hand, as $g_X$ becomes smaller, the relic density of the DM candidates increases considerably and tends to overclose the Universe. For example, the small value of $g_X$ from BP5 in Table \ref{table:benchmarkpoints} leads to $\Omega_X h^2 \simeq 6.2$, in which case $X^{1,2}$ are also completely depleted and $X^{3'}$ makes up $100\%$ of the relic density. Furthermore, the dependence of $M_{h_2}$ on $g_X$ means that there can be only two resonant dips, corresponding to $M_{h_1}/2$ and $M_{h_3}/2$. This is in contrast to the non-CSI version of the model \cite{Gross2015} where there should be three resonant dips, corresponding to $M_{h_1}/2$, $M_{h_2}/2$, $M_{h_3}/2$, since in that case $M_{h_2}$ does not depend on $g_X$. As a result, the CSI version of the model that we consider is in general more constrained.

%%%%%%%%%%%%%%%%%%%%%%%%%%%%%%%%%%%%
\subsubsection{Case $v_1 \simeq v_2$}
\begin{figure}
\includegraphics[width=.5\textwidth]{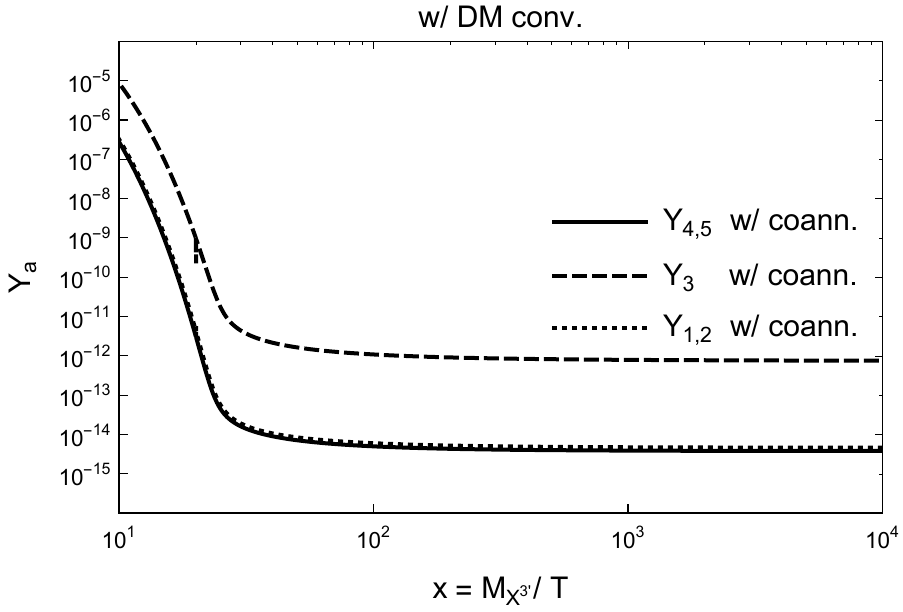}
\includegraphics[width=.5\textwidth]{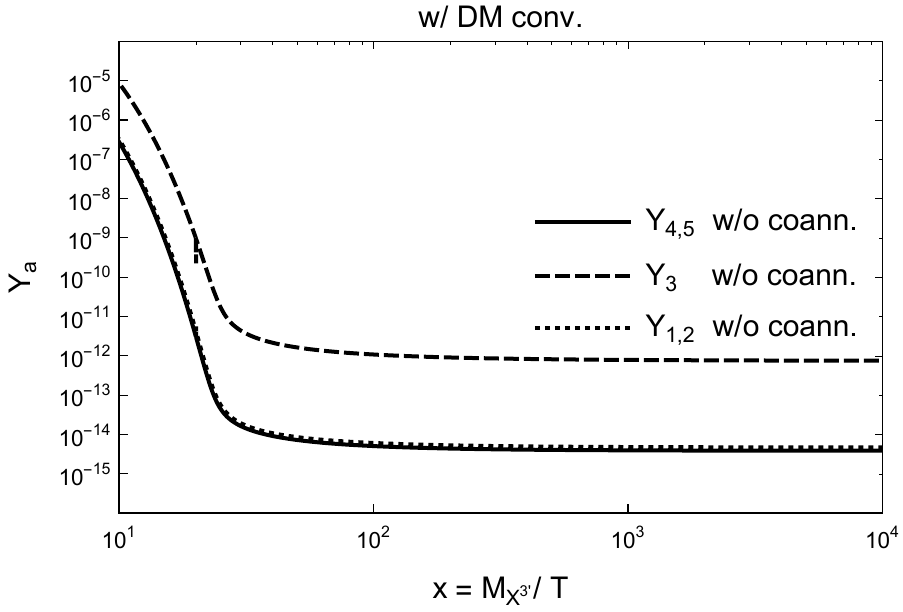}
\caption{The yields $Y_{1,2}$, $Y_3$, and $Y_{4,5}$ in terms of $x = M_{X^{3'}}/T$ for BP2. The right plot has been obtained neglecting the coannihilation terms in the Boltzmann equations. These terms are included in the left plot. The difference is very small since most $X^{1,2}$ and $X^{4,5}$ particles are converted to $X^{3'}$.}
\label{fig:YieldPlotsBP2a}
\end{figure}

%%%%%%%%%%%%%%%%%%%%%%%%%%%%%%%%%%%%
In this case, $X^{3'}$ is nearly $20\%$ lighter than $X^1$ and $X^2$ [cf. \eqref{eq:secondlimitmasses}] while $X^4$ and $X^5$ are almost degenerate with the latter ones. Therefore, coannihilations between $X^{1,2}$ and $X^{4,5}$ may occur around the time of freeze-out and influence the relic density of these four particles. Since the semiannihilations $X^{1,2} X^{3'} \rightarrow X^{2,1} h_i$ are now phase-space suppressed, the Boltzmann equations governing the number densities of $X^{1,2}$ and $X^{4,5}$ are almost identical. We therefore expect their relic number densities to be very close. This is indeed the case as can be seen in Fig. \ref{fig:YieldPlotsBP2a}. 

There, we also distinguish between the cases when coannihilations are switched on (left) and switched off (right). The effect is clearly insignificant, and in both cases the DM candidates $X^1$, $X^2$, and $X^{3'}$ comprise approximately $1\%$, $1\%$, and $98\%$ of the total relic density, respectively. The dominant phenomenon is DM conversion since most $X^{1,2}$ and $X^{4,5}$ have had enough time to annihilate to $X^{3'}$. 

\begin{figure}
\includegraphics[width=.5\textwidth]{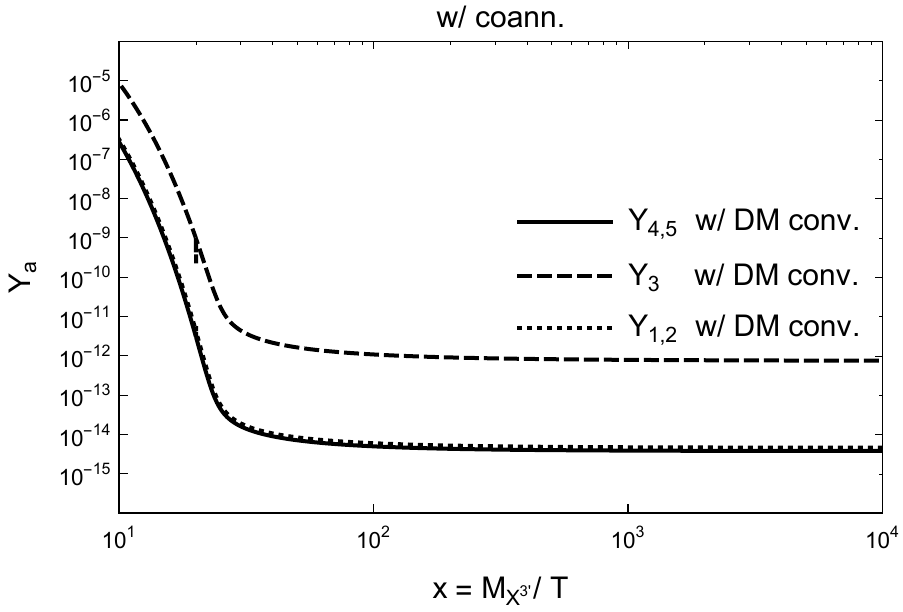}
\includegraphics[width=.5\textwidth]{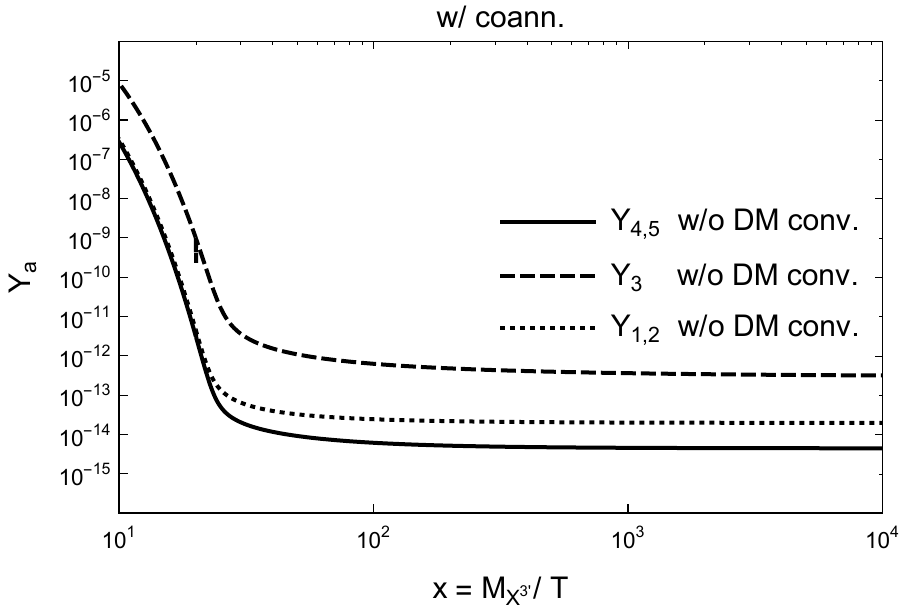}
\caption{The yields $Y_{1,2}$, $Y_3$, and $Y_{4,5}$ in terms of $x = M_{X^{3'}}/T$ for BP2. The right plot has been obtained neglecting the DM conversion terms in the Boltzmann equations. These terms are included in the left plot. The DM conversion processes are significant since many $X^{1,2}$ particles are converted to $X^{3'}$.}
\label{fig:YieldPlotsBP2b}
\end{figure}

We display the importance of this effect in Fig. \ref{fig:YieldPlotsBP2b}, where coannihilations are switched on, but this time we distinguish between the cases when DM conversion is switched on (left) and off (right). With DM conversion switched on, the DM candidates $X^1$, $X^2$, and $X^{3'}$ comprise again $1\%$, $1\%$, and $98\%$ of the total relic density. With DM conversion switched off, $X^1$, $X^2$, and $X^{3'}$ comprise around $7\%$, $7\%$, and $86\%$ of the total relic density, respectively. Moreover, the total relic density is almost $2$ times larger in the former case (DM conversion on) than in the latter case (DM conversion off). This can be attributed to the fact that without DM conversion freeze-out is delayed and more DM particles have time to annihilate to SM particles.

%%%%%%%%%%%%%%%%%%%%%%%%
\subsection{Direct detection}
\label{subsec:DD}
%%%%%%%%%%%%%%%%%%%%%%%%
Maybe the best prospect for validating the WIMP DM paradigm is through the direct detection of DM particles at deep underground facilities. Many experiments are in progress, and hopefully we may soon get a glimpse of this dark world.

Interactions between the DM particles $X_\mu^{1,2,3'}$ and the nucleons $N$ can be mediated through a $t$-channel exchange of the scalar bosons $h_i$. For the individual DM components, the corresponding spin-independent elastic scattering cross sections are
\be 
\sigma_{1,2}^{\rm SI}  =  \frac{f_N^2}{16 \pi v_h^2} \frac{m^4_N}{\left( M_{X^{1,2}} + m_N \right)^2} \left| g^2_X v_2 \sum^3_{i=1} \frac{\mathcal{R}_{i3} \mathcal{R}_{1i}}{M^2_{h_i}} \right|^2 ,
\ee
\be
\begin{split}
\sigma_{3}^{\rm SI} & =  \frac{f_N^2}{16 \pi v_h^2} \frac{m^4_N}{\left( M_{X^{3'}} + m_N \right)^2} \left| \frac{4}{3} g^2_X v_1 \sin^2\delta \sum^3_{i=1} \frac{\mathcal{R}_{i2} \mathcal{R}_{1i}}{M^2_{h_i}} \right. \\
               & \left. \qquad \qquad \qquad \qquad \qquad \quad + \frac{1}{3} g_X^2 v_2 \left( \cos 2\delta + 2 - \sqrt{3} \sin 2 \delta \right) \sum^3_{i=1} \frac{\mathcal{R}_{i3} \mathcal{R}_{1i}}{M^2_{h_i}} \right|^2,
\end{split}
\ee
where $f_N \simeq 0.3$~\cite{DiChiara2015, Cline2013, Alarcon2012, Junnarkar2013, Alarcon2014, Crivellin2014, Hoferichter2015} is the nucleon form factor and $m_N = 0.939\GeV$ is the average nucleon mass.

Since we have three DM candidates with different masses ($M_{X^1} = M_{X^2} > M_{X^{3'}}$), not all of them contribute equally to the local DM density which in direct detection experiments is assumed to be composed of a single DM species. Nevertheless, we may assume that the contribution of each DM species to the local density is equal to the contribution of that particular species to the relic density and consequently construct the effective cross sections~\cite{Aoki2012, Aoki2014, Esch2014}
\be 
\sigma_a^{\rm eff} = \sigma_a^{\rm SI} \left( \frac{\Omega_{X^a} h^2}{\Omega_X h^2} \right).
\ee

For example, BP3 in Table \ref{table:benchmarkpoints} reproduces the observed DM relic density within $3 \sigma$, with $X^1$, $X^2$, and $X^{3'}$ comprising approximately $5\%$, $5\%$, and $90\%$ of its total. The resulting effective cross sections are then

\begin{eqnarray*}
\sigma^{\rm eff}_{1,2} &=& 1.46715 \times 10^{-47} \,\, \rm cm^2, \\ 
\sigma^{\rm eff}_{3} &=& 2.77662 \times 10^{-46} \,\, \rm cm^2.
\end{eqnarray*}
Both of these numbers are well below the limits set by the LUX experiment~\cite{Akerib2014a}, but are nevertheless within the reach of future experiments such as LZ~\cite{Malling2011} and XENON1T~\cite{Aprile:2015uzo}.
%%%%%%%%%%%%%%%%%%%%%%%%
\section{Summary and conclusions}
\label{sec:conclusions}
%%%%%%%%%%%%%%%%%%%%%%%%
In the present article we have examined a classically scale-invariant extension of the SM, enlarged by a weakly coupled dark $SU(3)_X$ gauge group. The extra sector consists of the eight dark gauge bosons and two complex scalar triplets. Under mild assumptions on the parameters of the scalar potential of the model the scalar triplets can develop nonvanishing VEVs and break the extra $SU(3)_X$ completely via the Coleman-Weinberg mechanism. Eight of the $12$ scalar degrees of freedom are absorbed by the dark gauge bosons, rendering them all massive. We focused on and analyzed the case in which the symmetry breaking pattern involves two VEVs. As a result of the portal couplings of the dark scalars to the Higgs field, the dark gauge symmetry breakdown triggers electroweak symmetry breaking. In the framework of the Gildener-Weinberg formalism we considered the full one-loop effective potential. At one-loop level the pseudo--Nambu-Goldstone boson of broken classical scale symmetry receives a large radiative mass. Out of the massive dark gauge bosons the lightest three of them are almost degenerate in mass and also stable due to an intrinsic $Z_2 \times Z'_2$ discrete symmetry of $SU(3)_X$. These are identified as DM candidates.

The parameters of the model and the mass patterns resulting from symmetry breaking have subsequently been subjected to the various existing theoretical and experimental constraints. The requirements on the tree-level and one-loop effective scalar potential to be bounded from below have been analyzed. Constraints arising from LHC searches and measurements of the electroweak parameters $S$ and $T$ have also been examined. Thus, we obtained five benchmark points for the parameters of the model that stabilize the vacuum, satisfy the experimental constraints, and reproduce the measured mass for the observed Higgs boson. 

Having analyzed the phenomenological viability of the model, a comprehensive DM analysis was undertaken. After identifying the relevant DM processes (annihilations, semiannihilations, coannihilations, and DM conversions), the set of coupled Boltzmann equations was constructed,  describing the number density evolution of the DM candidates in order to obtain their total relic density and compare it to the measured value. The Boltzmann equations were solved numerically in two cases defined by the VEVs of the $SU(3)_X$ scalar fields. 

In the first case, the VEV separation was large ($v^2_1 \gg v_2^2$) and the three dark gauge boson candidates $X^1$, $X^2$, and $X^{3'}$ were nearly degenerate in mass. This case may seem similar to the dark $SU(2)_X$ model (recently considered in Refs. \cite{Karam2015} and \cite{Hambye2013, Carone2013, Khoze2014, Khoze2016}) where the extra gauge symmetry gets broken by a complex scalar doublet. There, the three dark gauge bosons are completely degenerate in mass and contribute equally to the DM relic density. In the $SU(3)_X$ model, however, even though $X^{1,2,3'}$ are nearly degenerate in mass, the lightest of the three ($X^{3'}$) is the predominant DM component. This occurs mainly due to the mixing between $X^{3'}$ and $X^{8'}$ which means that more Feynman diagrams contribute to the semiannihilation processes $X^{1,2} X^{1,2} \rightarrow X^{3'} h_i$ and the annihilation processes $X^{3'} X^{3'} \rightarrow h_i h_j$. Also, even though the mass splitting is small, some of the $X^{1,2}$ particles are converted to $X^{3'}$ and increase its final relic density. Finally, as it is transparent in the framework of the GW formalism employed, the pNGB mass depends on all the other masses of the model. Consequently, there can be only one resonant dip for the DM relic density in the $SU(2)_X$ model (corresponding to $M_{\rm Higgs}/2$) and two in the $SU(3)_X$ model (corresponding to $M_{h_1}/2$ and $M_{h_3}/2$). Therefore, in general, enlarging the gauge group means that more scalars are needed in order to break it, which leads to a larger parameter space that may be compatible with cosmological observations.

In the second case, the VEVs were very close ($v_1 \simeq v_2$). This resulted in $X^{3'}$ being around $20\%$ lighter than $X^1$, $X^2$ (which were exactly degenerate) and $X^4$, $X^5$ (which were exactly degenerate too) now being close in mass with $X^1$ and $X^2$. Therefore, possible coannihilation effects had to be examined. Nevertheless, it turned out that the dominant process was DM conversion and $X^{3'}$ was again the predominant DM component. Finally, we determined that the DM candidates have viable prospects of being directly detected by future underground experiments.

%%%%%%%%%%%%%%%%%%%%%%%%
\section*{Acknowledgements}
A.K. would like to thank Dimitrios Karamitros for useful discussions. K.T. thanks the CERN Theory Division for its hospitality during the period when this work was completed.
%%%%%%%%%%%%%%%%%%%%%%%%
\begin{appendices}
\section{Oblique Parameters}
\label{app:obliques}
%%%%%%%%%%%%%%%%%%%%%%%%%
The $S$ and $T$ parameters are given in this model by the expressions (see also~\cite{Farzinnia2013, Bian2015, Hambye2009})

\begin{eqnarray}
\label{eq:ST}
S &=& \frac{1}{24\pi}\Biggl\{ \mathcal{R}_{11}^{2}\left[\log R_{h_{1}h}
+G(M_{h_{1}}^{2},M_{Z}^{2})-G(M_{h}^{2},M_{Z}^{2})\right]\nonumber\\
&&\qquad +\mathcal{R}_{12}^{2}\left[\log R_{h_{2}h}
+G(M_{h_{2}}^{2},M_{Z}^{2})-G(M_{h}^{2},M_{Z}^{2})\right]\nonumber\\
&&\qquad +\mathcal{R}_{13}^{2}\left[\log R_{h_{3}h}
+G(M_{h_{3}}^{2},M_{Z}^{2})-G(M_{h}^{2},M_{Z}^{2})\right]\Biggr\},\\
T &=& \frac{3}{16\pi\sin^{2}\theta_{W}}\Biggl\{ \mathcal{R}_{11}^{2}\left[\frac{1}{\cos^{2}\theta_{W}}
\left(\frac{\log R_{Zh_{1}}}{1-R_{Zh_{1}}}
-\frac{\log R_{Zh}}{1-R_{Zh}}\right)
-\left(\frac{\log R_{Wh_{1}}}{1-R_{Wh_{1}}}
-\frac{\log R_{Wh}}{1-R_{Wh}}\right)\right]\nonumber\\
&&\quad\quad+\mathcal{R}_{12}^{2}\left[\frac{1}{\cos^{2}\theta_{W}}
\left(\frac{\log R_{Zh_{2}}}{1-R_{Zh_{2}}}
-\frac{\log R_{Zh}}{1-R_{Zh}}\right)
-\left(\frac{\log R_{Wh_{2}}}{1-R_{Wh_{2}}}
-\frac{\log R_{Wh}}{1-R_{Wh}}\right)\right]\nonumber\\
&&\quad\quad +\mathcal{R}_{13}^{2}\left[\frac{1}{\cos^{2}\theta_{W}}
\left(\frac{\log R_{Zh_{3}}}{1-R_{Zh_{3}}}
-\frac{\log R_{Zh}}{1-R_{Zh}}\right)
-\left(\frac{\log R_{Wh_{3}}}{1-R_{Wh_{3}}}
-\frac{\log R_{Wh}}{1-R_{Wh}}\right)\right]\Biggr\},
\end{eqnarray}

%%%%%%%%%%%%%%%%%%%%%%%%%%%%%%%%%%%%%%%%%%%
where the functions $R_{AB}$, $G(m_{A}^{2},m_{B}^{2})$, and $f(R_{AB})$ are given by
%%%%%%%%%%%%%%%%%%%%%%%%%%%%%%%%%%%%%%%%%%%

\begin{eqnarray}
R_{AB} &=& \frac{M_{A}^{2}}{M_{B}^{2}}\; ,
\end{eqnarray}
\begin{eqnarray}
G(M_{A}^{2},M_{B}^{2}) &=& - \frac{79}{3}+9R_{AB}-2R_{AB}^{2}
+(12-4R_{AB}+R_{AB}^{2})f(R_{AB})\nonumber\\
&&+(-10+18R_{AB}-6R_{AB}^{2}+R_{AB}^{3}-9\frac{R_{AB}+1}{R_{AB}-1})\log R_{AB},\\
f(R_{AB}) &=& 
\begin{cases}
\begin{array}{cc}
\sqrt{R_{AB}(R_{AB}-4)}\log\left|\frac{R_{AB}-2-\sqrt{R_{AB}(R_{AB}-4)}}{2}\right| 
& R_{AB}>4,\\
0 & R_{AB}=4,\\
2\sqrt{R_{AB}(4-R_{AB})}\arctan\sqrt{\frac{4-R_{AB}}{R_{AB}}} & R_{AB}<4.
\end{array}
\end{cases}
\end{eqnarray}

%%%%%%%%%%%%%%%%%%%%%%%%%
\section{RGEs}
\label{app:RGEs}
%%%%%%%%%%%%%%%%%%%%%%%%
%%%%%%%%%%%%%%%%%%%%%%%

In this appendix, we present the two-loop gauge RGEs, as well as the one-loop RGEs for the Yukawa and scalar couplings. However, in our numerical analysis we used the full set of two-loop RGEs obtained using \texttt{SARAH}~\cite{Staub2014a,Staub2015}. Defining $\beta_\kappa \equiv \left( 4 \pi \right)^2 \frac{d \kappa}{d \ln \mu}$, the RGEs have the form

{\allowdisplaybreaks \begin{align} 
\beta_{g_1} & =  
\frac{41}{10} g_{1}^{3} + \frac{1}{(4\pi)^2} \frac{1}{50} g_{1}^{3} \Big( 199 g_{1}^{2} + 135 g_{2}^{2} + 440 g_{3}^{2} -85 y_t^2 \Big), \\ 
\beta_{g_2} & =  -\frac{19}{6} g_{2}^{3} + \frac{1}{(4\pi)^2} \frac{1}{30} g_{2}^{3} \Big( 27 g_{1}^{2} + 175 g_{2}^{2} + 360 g_{3}^{2} -45 y_t^2 \Big),  \\  
\beta_{g_3} & =  -7 g_{3}^{3}  + \frac{1}{(4\pi)^2}\frac{1}{10} g_{3}^{3} \Big( 11 g_{1}^{2} + 45 g_{2}^{2} - 260 g_{3}^{2} - 20 y_t^2  \Big),\\  
\beta_{g_X} & =  -\frac{32}{3} g_{X}^{3}  -\frac{1}{(4\pi)^2}\frac{284}{3} g_{X}^{5}, \\
\beta_{y_t} & = y_t \left( \frac{9}{2} y_t^2 -\frac{17}{20} g_{1}^{2} -\frac{9}{4} g_{2}^{2} -  8 g_{3}^{2} \right), \\ 
\beta_{\lambda_h} & =  -6 y_t^4 +24 \lambda_{h}^{2}  +\lambda_h \left( 12 y_t^2 -\frac{9}{5} g_{1}^{2}  -9 g_{2}^{2} \right) +\frac{27}{200} g_{1}^{4} +\frac{9}{20} g_{1}^{2} g_{2}^{2} +\frac{9}{8} g_{2}^{4}  +3 \lambda_{h1}^{2} +3 \lambda_{h2}^{2},  \\ 
\beta_{\lambda_{1}} & =  
-16 g_{X}^{2} \lambda_{1}  + 28 \lambda_{1}^{2}  -2 \lambda_{3} \lambda_{4}  + 2 \lambda_{h1}^{2}  + 3 \lambda_{3}^{2}  + \frac{13}{6} g_{X}^{4}  + \lambda_{4}^{2} + \lambda_{5}^{2},\\ 
\beta_{\lambda_{2}} & =  
-16 g_{X}^{2} \lambda_{2}  + 28 \lambda_{2}^{2}  -2 \lambda_{3} \lambda_{4}  + 2 \lambda_{h2}^{2}  + 3 \lambda_{3}^{2}  + \frac{13}{6} g_{X}^{4}  + \lambda_{4}^{2} + \lambda_{5}^{2},\\ 
\beta_{\lambda_{h1}} & = \lambda_{h1} \left( -\frac{9}{10} g_{1}^{2}  -\frac{9}{2} g_{2}^{2}  -8 g_{X}^{2} +12 \lambda_h  -4 \lambda_{h1} +16  \lambda_{1} +6 y_t^2  \right) +6 \lambda_{h2} \lambda_{3} -2 \lambda_{h2} \lambda_{4}, \\ 
\beta_{\lambda_{h2}} & =  \lambda_{h2} \left( -\frac{9}{10} g_{1}^{2}  -\frac{9}{2} g_{2}^{2}  -8 g_{X}^{2} +12 \lambda_h  -4 \lambda_{h2} +16  \lambda_{1} +6 y_t^2  \right) +6 \lambda_{h1} \lambda_{3} -2 \lambda_{h1} \lambda_{4}, \\ 
\beta_{\lambda_{3}} & = \lambda_{3} \left( -16 g_{X}^{2}  + 16 \lambda_{2}  + 16 \lambda_{1} -4 \lambda_{3} \right) -2 \lambda_{4}^{2}  -2 \lambda_{5}^{2}    -4 \lambda_{1} \lambda_{4}  -4 \lambda_{2} \lambda_{4}  + 4 \lambda_{h1} \lambda_{h2}  -\frac{11}{6} g_{X}^{4}, \\
\beta_{\lambda_{4}} & =  
10 \lambda_{5}^{2}  -16 g_{X}^{2} \lambda_{4}  + 4 \lambda_{1} \lambda_{4}  + 4 \lambda_{2} \lambda_{4}  + 6 \lambda_{4}^{2}  -8 \lambda_{3} \lambda_{4}  + \frac{5}{2} g_{X}^{4}, \\ 
\beta_{\lambda_{5}} & = 4 \lambda_{5} \left(-2 \lambda_{3}  -4 g_{X}^{2}  + 4 \lambda_{4}  + \lambda_{1} + \lambda_{2}\right).
\end{align} }
%%%%%%%%%%%%%%%%%%%%%%%

For the SM gauge couplings and the top quark Yukawa coupling we specify the boundary conditions at the top quark pole mass $M_t$~\cite{Oda2015b,Buttazzo2013},

%%%%%%%%%%%%%%%%%%%%%%%
\begin{align}
g_1(\mu=M_t) &= \sqrt{\frac{5}{3}} \left( 0.35830 + 0.00011 \left( \frac{M_t}{\GeV} - 173.34 \right) 
			- 0.00020\left(  \frac{M_W - 80.384{\GeV}}{0.014{\GeV}} \right) \right), \\
g_2(\mu=M_t) &= 0.64779 + 0.00004 \left( \frac{M_t}{\GeV} - 173.34 \right) 
			+ 0.00011 \left( \frac{M_W - 80.384{\GeV}}{0.014{\GeV}} \right),  \\			
g_3(\mu=M_t) &= 1.1666 + 0.00314 \left( \frac{\alpha_s(M_Z) - 0.1184}{0.0007} \right) 
			-  0.00046 \left( \frac{M_t}{\GeV} - 173.34 \right),  \\
y_t(\mu=M_t) &= 0.93690 + 0.00556 \left( \frac{M_t}{\GeV} - 173.34 \right) 
			- 0.00042 \left( \frac{\alpha_s(M_Z) - 0.1184}{0.0007} \right), 
\end{align}
%%%%%%%%%%%%%%%%%%%%%%%
whereas the dark gauge coupling is defined at the scale of the lightest dark gauge boson $g_X(M_{X^{3'}})$. The boundary conditions for the scalar couplings are specified at the renormalization scale $\Lambda$ where the tree-level potential is minimized.
%%%%%%%%%%%%%%%%%%%%%%%
\section{Kinematics}
\label{app:kin}
The expressions for the Mandelstam variables $s$, $t$, $u$ in the center of mass (CM) frame for the general process $a+b\rightarrow c+d$ are
\ba 
s &=& \left( E_a + E_b \right)^2, \label{eq:MandelS} \\
t &=& m_a^2 + m_c^2 - 2 E_a E_c + 2 p_a(s) p_c(s) \cos \theta , \\
u &=& m_a^2 + m_d^2 - 2 E_a E_d - 2 p_a(s) p_d(s) \cos \theta ,
\ea
where $\theta$ is the CM scattering angle and $p_i(s)=\vert \vec{p}_i \vert$. The energies $E_a$, $E_b$, $E_c$, $E_d$ and the 3-momenta $p_a$, $p_b$, $p_c$, $p_d$ can be expressed in terms of the CM energy squared $s$ as
\be
\begin{array}{l} 
E_a =\frac{1}{2\sqrt{s}} \left( s + m^2_a - m^2_b \right),\,\,\,E_b = \frac{1}{2\sqrt{s}} \left( s + m^2_b - m^2_a \right), \\
\,\\
E_c = \frac{1}{2\sqrt{s}} \left( s + m^2_c - m^2_d \right),\,\,\,E_d = \frac{1}{2\sqrt{s}} \left( s + m^2_d - m^2_c \right), 
\end{array}
\ee
\be
\begin{array}{l}
p_{\rm in}(s) \equiv p_a(s) = p_b(s) = \left[\frac{s}{4}-\frac{1}{2}(m_a^2+m_b^2)+\frac{1}{4s}(m_a^2-m_b^2)^2\right]^{1/2} 
\, ,\\
p_{\rm out}(s) \equiv p_c(s) = p_d(s) = \left[\frac{s}{4}-\frac{1}{2}(m_c^2+m_d^2)+\frac{1}{4s}(m_c^2-m_d^2)^2\right]^{1/2}\,.
\end{array}
\ee
Using $s+t+u=m^2_a+m^2_b+m^2_c+m^2_d$ we can write
\ba 
t &=& \frac{1}{2} \left[ m^2_a+m^2_b+m^2_c+m^2_d - s + (t - u) \right], \\
u &=& \frac{1}{2} \left[ m^2_a+m^2_b+m^2_c+m^2_d - s - (t - u) \right],
\label{eq:MandelU}
\ea
and express $t-u$ as
\be 
t - u = - \frac{1}{s}(m^2_a - m^2_b)(m^2_c - m^2_d) + 4 p_a(s) p_c(s) \cos\theta.
\label{eq:TminusU}
\ee
In view of the above, any function $f(s,t,u)$ is a function of $s$ and the incoming momentum projection $p_{\rm in}(s)\cos\theta$. Finally, the relative velocity is
\be v_r\,=\,|\vec{\beta}_a-\vec{\beta}_b|=\frac{p_{\rm in}\sqrt{s}}{E_a E_b}\,.\ee

\medskip

\medskip

\end{appendices}

%%%%%%%%%%%%%%%%%%%%%%%
%%%%%%%%%%%%%%%%%%%%%%%
%\printbibliography
%\addbibresource{References.bib}
%,C:/Users/ALEX/Dropbox/2ndPaper/CSI-models
\bibliography{C:/Users/ALEX/Dropbox/2ndPaper/References}{}
\bibliographystyle{utphys}

\end{document}